\documentclass[useAMS,usenatbib]{mn2e}

\usepackage{graphicx}
\usepackage{amssymb,amsmath}
\usepackage{txfonts}
\usepackage[draft]{hyperref}
\usepackage{pifont}
\usepackage{multirow}
\usepackage{hhline}

\def\Msun{ M_\odot}
\def\Rsun{ R_\odot}
\def\Lsun{ L_\odot}

\def\Mp{M_{\rm p}}
\def\Rp{R_{\rm p}}

\def\rf{r_\textrm{F}}

\def\Ms{M_{\star}}
\def\Rs{R_{\star}}
\def\Ls{L_{\star}}

\def\Mbd{M_{{\rm BD}}}

\def\Mearth{ M_\oplus}
\def\Rearth{ R_\oplus}
\def\Searth{ S_\oplus}

\def\G{\mathcal{G}}

\def\d{\mathrm{d}}

\newcommand{\Tpp}{TRAPPIST-1}

\usepackage[normalem]{ulem}
\usepackage{color}
\definecolor{blue}{RGB}{0,0,255}
\definecolor{red}{RGB}{255,0,0}
\definecolor{green}{RGB}{0,200,0}
\definecolor{black}{RGB}{0,0,0}

\title[Water loss from terrestrial planets orbiting ultracool dwarfs]{Water loss from terrestrial planets orbiting ultracool dwarfs: \\
Implications for the planets of TRAPPIST-1}
\author[E. Bolmont, F. Selsis, J. E. Owen, I. Ribas, S. N. Raymond, J. Leconte and M. Gillon]{E. Bolmont$^{1}$\thanks{E-mail:
emeline.bolmont@unamur.be}, F. Selsis$^{2,3}$, J. E. Owen$^4$\thanks{Hubble Fellow}, I. Ribas$^{5}$, S. N.
Raymond$^{2,3}$, J. Leconte$^{2,3}$, \newauthor \& M. Gillon$^{6}$ \\
$^{1}$ NaXys, Department of Mathematics, University of Namur, 8 Rempart de la Vierge, 5000 Namur, Belgium \\
$^{2}$ Univ. Bordeaux, LAB, UMR 5804, F-33615, Pessac, France\\
$^{3}$ CNRS, LAB, UMR 5804, F-33615, Pessac, France \\
$^{4}$ Institute for Advanced Study, Einstein Drive, Princeton NJ 08540, USA \\
$^{5}$ Institut de Ci\`encies de l'Espai (CSIC-IEEC), Carrer de Can Magrans s/n, Campus UAB, 08193 Bellaterra, Spain \\
$^{6}$ Institut d'Astrophysique et de G\'eophysique, Universit\'e de Li\`ege, All\'ee du 6 Ao\^ut 19C, 4000 Li\`ege, Belgium} 
\begin{document}

\date{Accepted xx xx xx. Received xx xx xx; in original form xx xx xx}

\pagerange{\pageref{firstpage}--\pageref{lastpage}} \pubyear{xx}

\maketitle

\label{firstpage}

\begin{abstract}

Ultracool dwarfs (UCD; $T_{\rm eff}<\sim3000~$K) cool to settle on the main sequence after $\sim$1 Gyr. 
For brown dwarfs, this cooling never stops. 
Their habitable zone (HZ) thus sweeps inward at least during the first Gyr of their lives.
Assuming they possess water, planets found in the HZ of UCDs have experienced a runaway greenhouse phase too hot for liquid water prior to entering the HZ. 
It has been proposed that such planets are desiccated by this hot early phase and enter the HZ as dry worlds. 
Here we model the water loss during this pre-HZ hot phase taking into account recent upper limits on the XUV emission of UCDs and using 1D radiation-hydrodynamic simulations. 
We address the whole range of UCDs but also focus on the planets recently found around the $0.08~\Msun$ dwarf TRAPPIST-1.

Despite assumptions maximizing the FUV-photolysis of water and the XUV-driven escape of hydrogen, we find that planets can retain significant amounts of water in the HZ of UCDs, with a sweet spot in the $0.04$~--~$0.06~\Msun$  range. 
We also studied the TRAPPIST-1 system using observed constraints on the XUV-flux.
We find that TRAPPIST-1b and c may have lost as much as 15 Earth Oceans and planet d -- which might be inside the HZ -- may have lost less than 1 Earth Ocean. 
Depending on their initial water contents, they could have enough water to remain habitable. 
TRAPPIST-1 planets are key targets for atmospheric characterization and could provide strong constraints on the water erosion around UCDs.

\end{abstract}


\section{Introduction}
Earth-like planets have been detected in the HZs \citep[defined in][]{Kasting1993, Selsis2007a, Kopparapu2013} of early type M-dwarfs \citep[e.g.,][]{Quintana2014}.
Here we address the potential habitability of planets orbiting even less massive objects: ultracool dwarfs (UCDs), which encompass brown dwarfs (BDs) and late type M-dwarfs. 
The first gas giant was detected around a BD by~\cite{Han2013}.
Very recently, \citet{Gillon2016} discovered 3 Earth-sized planets close to an object of mass $0.08~\Msun$ (just above the theoretical limit of $\Ms \sim 0.075~\Msun$ between brown dwarfs and M-dwarfs, \citealt{ChabrierBaraffe1997}).
The two inner planets in this system have insolations between 4.25 and 2.26 times Earth's.
The orbital period of planet d is still undetermined but between 4.5 and 73 days. 
The planet receives a stellar flux in the range 0.02-1 times the Earth one, which includes a large fraction of the habitable zone of Trappist-1 (0.023~au -- 0.048~au, \citealt{Kopparapu2013}).
The atmospheres of these planets could be probed with facilities such as the HST and JWST, which makes them all the more interesting to study \citep{BarstowIrwin2016}.

BDs (i.e., objects of mass $0.01 \leq \Ms/\Msun \leq 0.07$) do not fuse hydrogen in their cores \citep{Spiegel2011}. 
They contract and become fainter in time. 
Their HZs therefore move inward in concert with their decreasing luminosities~\citep{Andreeshchev2004, Bolmont2011}. 
Nonetheless, for BDs more massive than $0.04~\Msun$ a planet on a fixed or slowly-evolving orbit can stay in the HZ for Gyr timescales \citep{Bolmont2011}. 
In this study we consider UCDs of mass up to $0.08~\Msun$.
A UCD of mass $0.08~\Msun$ is actually a late-type M-dwarf.
As for a BD, its luminosity also decreases with time but its mass is high enough so that it starts the fusion of hydrogen in its core.
From that moment on, the HZ stops shrinking, allowing close-in planets to stay more than 10~Gyr in the HZ.
However, any planet that enters the HZ has spent time in a region that is too hot for liquid water. 
A planet experiencing a runaway greenhouse around a Sun-like star is expected to lose considerable amounts of water (one Earth ocean in less than one Gyr). 
This is due to H$_2$O photolysis by FUV photons and thermal escape of hydrogen due to XUV heating of the upper atmosphere. 
This is the current scenario to explain the water depletion in the atmosphere of Venus and its high enrichment in deuterium \citep{Solomon1991}. 
Could planets in the HZs of UCDs, like Venus, have lost all of their water during this early hot phase?

\citet{BarnesHeller2013} found that planets entering the BD habitable zone are completely dried out by the hot early phase. 
They concluded that BDs are unlikely candidates for habitable planets. 
Here we present a study of water loss using more recent estimates for the X-ray luminosity of very low mass stars and confronting results obtained with 1D radiation-hydrodynamic mass-loss simulations.
We find that -- even in a standard case scenario for water retention -- a significant fraction of an initial water reservoir (equivalent to one Earth ocean, $M_{\rm H_2O}=1.3\times10^{21}$~kg) could still be present upon reaching the HZ.
Once planets enter the HZ, the water can condense onto the surface \citep[as is thought happened on Earth, once the accretion phase was over; e.g.][]{Matsui1986, Zahnle1988}.
Observation of these objects, such as the TRAPPIST-1 planets, would probably lift this uncertainty.

\section{Orbital evolution of planets in the ultracool dwarf habitable zone}

The UCD habitable zone is located very close-in, at just a few percent of an au~\citep[or less;][]{Bolmont2011}. 
Tidal evolution is therefore important. 
Due to UCDs' atmospheres low degree of ionization \citep{Mohanty2002} and the high densities, magnetic breaking is inefficient and cannot counteract spin-up due to contraction. 
UCDs' corotation radii -- where the mean motion matches the UCD spin rate -- move inward.  

\begin{figure}
	\begin{center}
	\includegraphics[width=\linewidth]{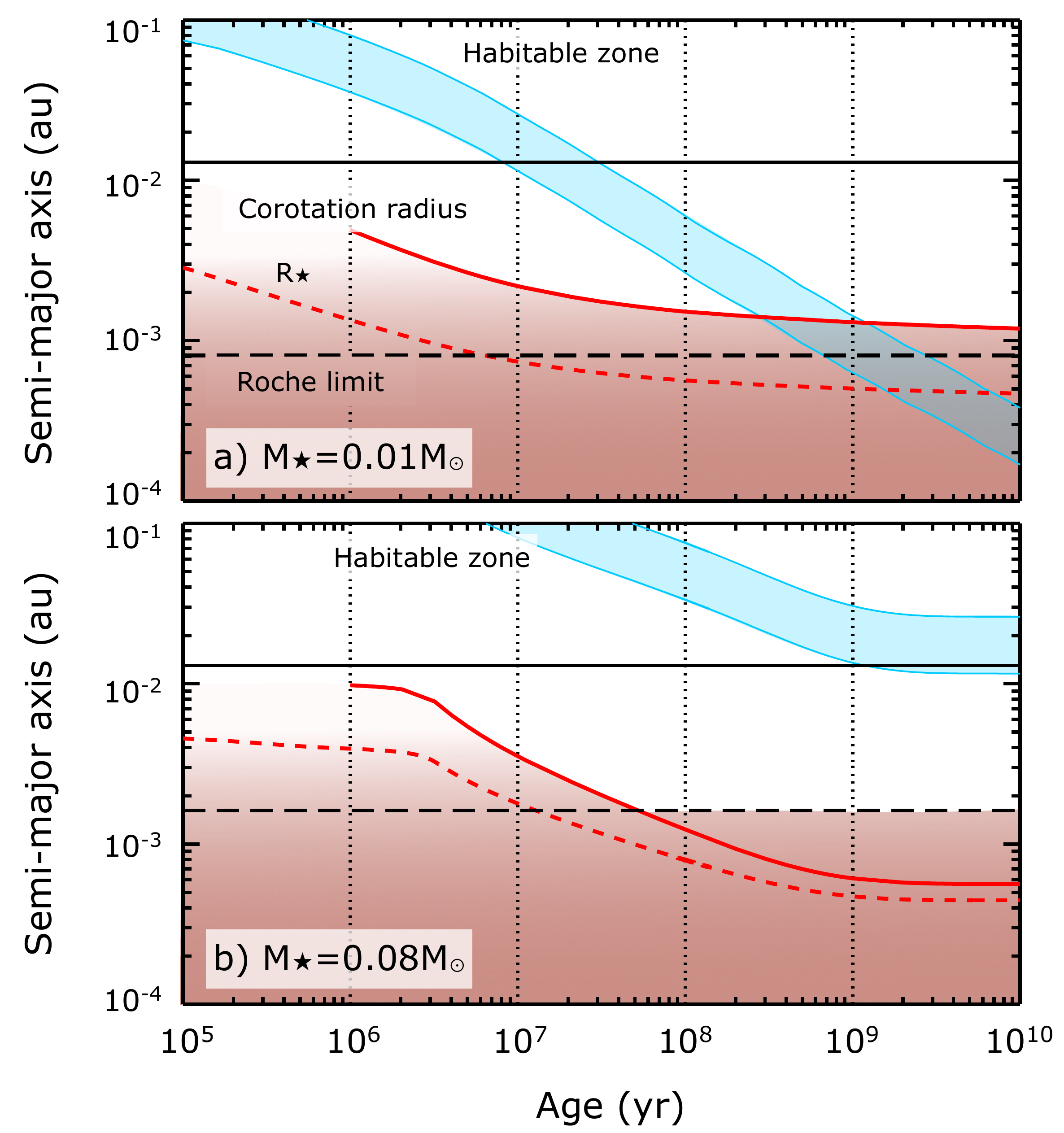}
	\caption{Evolution of the HZ limits for a UCD a) of $0.01~\Msun$ and b) of $0.08~\Msun$: the inner edge corresponds to a flux of $1.5$ that of the Earth (2049~W.m$^{-2}$) and the outer edge corresponds to a flux of 240~W.m$^{-2}$ \citep{Bolmont2011}. The full lines correspond to the corotation radius. The dashed lines correspond to the radius of the dwarf and the thick dashed lines correspond to the Roche limit (assuming an Earth-like planet). The horizontal full black line represents an orbit of 0.013~au.}
	\label{aM004_Mp1_eo001}
	\end{center}
	\end{figure}

Figure \ref{aM004_Mp1_eo001} shows the evolution of the HZ boundaries for two UCDs: one of mass $0.01~\Msun$ and one of mass $0.08~\Msun$. 
The time at which a planet reaches the HZ depends on its tidal orbital evolution as well as the cooling rate of the UCD.
The tidal evolution of a planet around a UCD is mainly controlled by its initial semi-major axis with respect to the corotation radius (in red and orange full lines in Figure \ref{aM004_Mp1_eo001}). 
A planet initially interior to the corotation radius migrates inwards due to the tide raised in the UCD and eventually falls on the UCD. 
A planet initially outside corotation migrates outward \citep{Bolmont2011}. 
There does exist a narrow region in which the moving corotation radius catches up to inward-migrating planets and reverses the direction of migration.  
The planet's probability of survival is smaller the farther inwards it is from the corotation radius, which is illustrated by the colored gradient area in Figure \ref{aM004_Mp1_eo001}.

Surviving planets cross the shrinking HZ at different times depending on their orbital distance. 
Close-in planets tend to stay longer in the HZ because the UCDs' luminosity evolution slows as it cools~\citep{ChabrierBaraffe1997}. 
\citet{Bolmont2011} showed that for UCDs of mass higher than $0.04~\Msun$, planets can spend up to several Gyr in the HZ.  
Although their earlier tidal histories vary, planets are on basically fixed orbits as they traverse the UCD habitable zone \citep{Bolmont2011}. 
We computed the mass loss taking into account the orbital tidal evolution, but we found that it has no effect on the result\footnote{In the worst case, i.e. if the UCD is very dissipative, this would mean that we underestimate the mass loss by $\sim1$\%.}. 
For the rest of this study we therefore consider the case of planets on fixed orbits. 
We consider planets from 0.005~au to $0.05$~au.
Figure \ref{aM004_Mp1_eo001} shows an orbit in this range (0.013~au), this orbit allows the planet to pass some time in the HZ of the $0.08~\Msun$ UCD. 
For a fixed initial rotation period of the UCD, planets around low mass UCDs could survive at smaller initial orbital radius, than around high mass UCDs. 

\section{Modeling water loss}\label{method}

\subsection{Energy-limited escape formalism}\label{enerlim}

In order to place the strongest possible constraints, we calculate the mass loss of the atmosphere via energy-limited escape \citep{Watson1981, Lammer2003}.
During a runaway phase, water is able to reach the stratosphere without condensing. 
Assuming photolysis is not a limiting process, water is photo-dissociated. 
In order to escape, water needs to reach the base of the hydrodynamic wind, which is much higher up. 
We here make the hypothesis that this is the case.
In this work, we do not consider the diffusion-limited escape, which would be responsible for a lower escape rate than the energy-limited escape rate.
The presence of other gases not considered here could bottleneck the diffusion of water vapor into the upper atmosphere.
The energy-limited escape mechanism requires two types of spectral radiation: FUV ($100$--$200$~nm) to photo-dissociate water molecules and XUV ($0.1$--$100$~nm) to heat up the exosphere. 
This heating causes atmospheric escape when the thermal velocity of atoms exceeds the escape velocity above the exobase. 
Non-thermal loss induced by stellar winds can in theory contribute significantly to the total atmospheric loss, either by the quiescent wind or by the coronal ejections associated with flares \citep{Lammer2007, Khodachenko2007}. 
For earlier type UCDs, this might be an issue, however we here consider UCDs of spectral type later than M7 type, for which there is no indication of winds that could enhance the atmospheric loss \citep[due to the low degree of ionization, see][]{Mohanty2002}.

Energy-limited escape considers that the energy of incident radiation with $\lambda < 100$~nm is converted into the gravitational energy of the lost atoms. 
This formalism was used in many studies \citep[e.g.,][]{BarnesHeller2013, LugerBarnes2015, HellerBarnes2015} but we use here the prescription of \citet{Selsis2007} linking the XUV flux $F_{{\rm XUV}}$(at d = $1$~au) to the mass loss rate $\dot{m}$:

\begin{equation}
\label{eqmassloss}
\epsilon \frac{F_{{\rm XUV}} \pi \Rp^2}{(a/1 {\rm au})^2}  = \frac{\G \Mp \dot{m}}{\Rp},
\end{equation}
where $\Rp$ is the planet's radius, $\Mp$ its mass and $a$ its semi-major axis. 
$\epsilon$ is the fraction of the incoming energy that is transferred into gravitational energy through the mass loss. 
This efficiency is not to be confused with the heating efficiency, which is the fraction of the incoming energy that is deposited in the form of heat, as only a fraction of the heat drives the hydrodynamic outflow (some being for instance conducted downward). 
This fraction has been estimated at about 0.1 by \citet{Yelle2004} and more recently at 0.1 or less in the limit of extreme loss by \citet{OwenWu2013}. 
The left term of Equation \ref{eqmassloss}, $F_{{\rm XUV}} \pi \Rp^2/(a/1 {\rm au})^2$, corresponds to the fraction of the XUV flux intercepted by the planet. 
Thus the mass loss rate is given by: 
\begin{equation}
\label{eqmassloss1}
\dot{m} = \epsilon \frac{F_{{\rm XUV}} \pi \Rp^3}{\G \Mp (a/1 {\rm au})^2}.
\end{equation}

Taking into account that the planet might undergo orbital migration or the evolution of the XUV flux the mass lost by the planet at a time $t$ is of:
\begin{equation}
\label{eqmassloss2}
m = \epsilon \frac{\pi \Rp^3}{\G \Mp}\int_0^t \frac{F_{{\rm XUV}}}{(a/1 {\rm au})^2} \d t.
\end{equation}


One could argue that the XUV cross section radius of the planet is larger than the planet's radius and is probably similar the Hill radius \citep[as in][which considered Jupiter-mass planets]{Erkaev2007}.
This assumption would mean that the water mass loss would be higher than what we propose to calculate here.
In a hydrodynamic outflow the effect of the Roche Lobe is to weaken the effect of gravity compared to the pressure force, its role is felt through the gradient of the potential. 
For transonic outflow the distance of the sonic point from the surface of the planet depends on the gradient of the potential, with shallow gradients (i.e. closer to the Hill radius) pushing the sonic point closer to the planet increases the mass loss.
We tested our hypothesis of taking the real radius of the planet with a hydrodynamical code \citep[see Section \ref{hydrocode} and][]{OwenAlvarez2016}. 
We find that, for the small rocky planets we consider here, this does not underestimate the mass loss as much as the work of \citet{Erkaev2007} would imply (taking the Roche radius), as the effect is of the order of a few percent.

\medskip

Venus probably lost its water reservoir by this mechanism. 
Radar observations of the Magellan satellite and ground-based observations showed that the last traces of water on Venus date back at least $1$~billion years \citep[this corresponds to the age of the surface;][]{Solomon1991}. 
Venus certainly experienced energy-limited hydrodynamic loss of water during several billion years \cite{Chassefiere2012}.
As UCDs are less bright than the Sun, we expect their XUV flux to be lower than the Sun's. 

In order to calculate the mass loss, we therefore need the XUV flux of the dwarfs and estimate the efficiency $\epsilon$. 
In the following, we give the observational constraints on the XUV luminosity of UCDs of spectral type later than M7 type.
We then use these values to compute the efficiency $\epsilon$ using 1D radiation-hydrodynamic mass-loss simulations.

\subsection{Physical inputs}\label{phys_input}

We first assume that protoplanetary disks around UCDs dissipate after 3~Myr \citep{Pascucci2009, PecautMamajek2016} ; this is our ``time zero''.
This assumption is a strong one, it is possible that disks live much longer than that, $\sim10$~Myr according to \citet{Pfalzner2014}. 
We consider that when the planet is embedded in the disk it is protected and does not experience mass loss.

Planets are thought to acquire water during and after accretion via impacts of volatile-rich objects condensed at larger orbital radii. 
On the one hand, if planets form in-situ in the HZs of UCDs then it may be difficult to retain water because of the rapid formation timescale and very high impact speeds \citep{Lissauer2007, Raymond2007}. 
On the other hand, because they form fast while embedded in a disk, they can acquire volatiles directly from the disk itself. 
It has been suggested that accreting H$_2$ this way can lead to the formation of H$_2$O \citep{IkomaGenda2006}.
In contrast, if these planets or some of their constituent planetary embryos migrated inward from wider orbital distances then they are likely to have significant water contents \citep{Raymond2008, Ogihara2009}. 
Although the origin of hot super-Earths is debated \citep[see][]{Raymond2008, Raymond2014}, several known close-in planets are consistent with having large volatile reservoirs \citep[e.g., GJ~1214 b;][]{Rogers2010, Berta2012}. 
We therefore consider it plausible for Earth-like planets to form in the UCD habitable zone with a water content at least comparable to Earth's surface reservoir and possibly much bigger.
For example, Earth contains several additional oceans of water trapped in the mantle~\citep{Marty2012}.
Water can also be trapped in the mantle of the planet during the formation process and perhaps released in the atmosphere at later times through volcanic activity. 

A key input into our model is the stellar flux at high energies. 
There are no observations of the EUV flux of UCDs. 
We assume here that the loss rate of water is not limited by the photo-dissociation of water. 
This assumes that FUV radiation is sufficient to dissociate all water molecules and produce hydrogen at a rate higher than its escape rate into space. 
X-ray observations with Chandra/ACIS-I2 exist for objects from M6.5 to L5 \citep[e.g.,][]{Berger2010, Williams2014} for the range $0.1$--$10$~keV ($0.1$--$12.4$~nm). 
This only represents a small portion of the XUV range considered in Equations \ref{eqmassloss1} and \ref{eqmassloss2}. 
For solar-type stars the flux in the whole XUV range is 2 to 5 times higher than in the X-ray range \citep{Ribas2005}. 
We thus multiply the value corresponding to the X-ray range by 5.
This constitute an upper limit of what one might expect, indeed for active UCDs the factor can be lower than 2 (e.g., for TRAPPIST-1 this factor is of 1.78, \citealt{Wheatley2016}).
We consider here that it can be used as a proxy for the whole XUV range.

On one hand, observations of objects of spectral types M0 to M7 show that the X-ray luminosity scales as $10^{-3}$ the bolometric luminosity $L_{{\rm bol}}$ \citep{Pizzolato2003}. 
On the other hand, more recent observations from \citet{Berger2010} and \citet{Williams2014} show that the X-ray luminosity of L dwarfs seems to scale as $10^{-5}\times$~$L_{{\rm bol}}$. 
However, some of these observations are actually non-detections, so the X-ray luminosity of objects like 2M0523-14 (L2--L3), 2M0036+18 (L3--L4) and 2M1507-16 (L5) must be even lower than $10^{-5}\times$~$L_{{\rm bol}}$. 
Between the populations of dwarfs of spectral type M0 to M7 and those of spectral type L0 to L6, there are dwarfs of spectral type M8 and M9 which do not really follow either trends.
TRAPPIST-1 belongs to this transition population.
That is why to study the water loss for the TRAPPIST planets we considered a wide range of XUV-luminosities that are representative of this transition region: from $L_{\rm X}/L_{{\rm bol}} < 10^{-5}$ \citep[as the analog dwarf VB~10][]{Williams2014} to $L_{\rm X}/L_{{\rm bol}} < 10^{-3.4}$ \citep[recent observations from][]{Wheatley2016}.
There is no indication of whether the X-ray luminosity varies in time.

For the UCDs of mass $0.01<\Ms/\Msun<0.08$, we consider two limiting cases for our mass loss calculation. 
In the first we adopt a value of $10^{-5}\times$~$L_{{\rm bol}}$. 
As the bolometric luminosity changes with time, we consider that the X-ray luminosity does as well. 
In the second case we assume that the X-ray luminosity does not vary with the bolometric luminosity but rather remains constant. 
We adopt the value of $10^{25.4}$~erg.s$^{-1} = 2.5\times10^{18}$~W from \citet{Williams2014}. 
This value corresponds to a X-ray detection (0.1--10~keV) of the object 2MASS13054019-2541059 AB of spectral type L2. 
To take into account the fact that the dwarfs with the higher masses considered in the range $0.01<\Ms/\Msun<0.08$ are possibly part of the transition population, we also tested two additional cases corresponding to a higher XUV emission: $L_{\rm X}/L_{{\rm bol}} < 10^{-4.5}$ and $L_{\rm X}=10^{26}$~erg.s$^{-1}$.
For each case we then use Equation \ref{eqmassloss2} to calculate the mass loss of a planet of $0.1~\Mearth$, $1~\Mearth$ and $5~\Mearth$.

We compared the XUV flux received by the planets of Figure \ref{aM004_Mp1_eo001} to the one Earth receives. 
Before reaching the HZ, they are at least a few times higher than Earth's incoming XUV flux. 


\subsection{Estimation of $\epsilon$ with a hydrodynamical code}\label{hydrocode}

In order to guide our calculations, we perform a set of 1D radiation-hydrodynamic mass-loss simulations based on the calculations of \citet{OwenAlvarez2016}. 
The simulations are similar in setup to those described by \citet{OwenAlvarez2016}, where we perform 1D spherically symmetric simulations using a modified version of the {\sc zeus} code \citep{Stone1992, Hayes2006}, along a streamline connecting the star and planet. 
We include tidal gravity, but neglect the effects of the Coriolis force as it is a small while the outflow remains sub-sonic \citep{MurrayClay2009, OwenJackson2012}. 
The radial grid is non-uniform and consists of 192 cells, the flow is evolved for 40 sound crossing times such that a steady state is achieved (which is checked by making sure the pseudo-Bernoulli potential and mass-flux are constant). 
We explicitly note that these simulations of a pure hydrogen atmosphere do not include line cooling from oxygen that maybe important in these flows, and as such these calculations should be considered as a maximum rate, as any other elements would lower the hydrogen loss by cooling the flow and colliding with the hydrogen atoms. 

\begin{figure}
	\begin{center}
	\includegraphics[width=\linewidth]{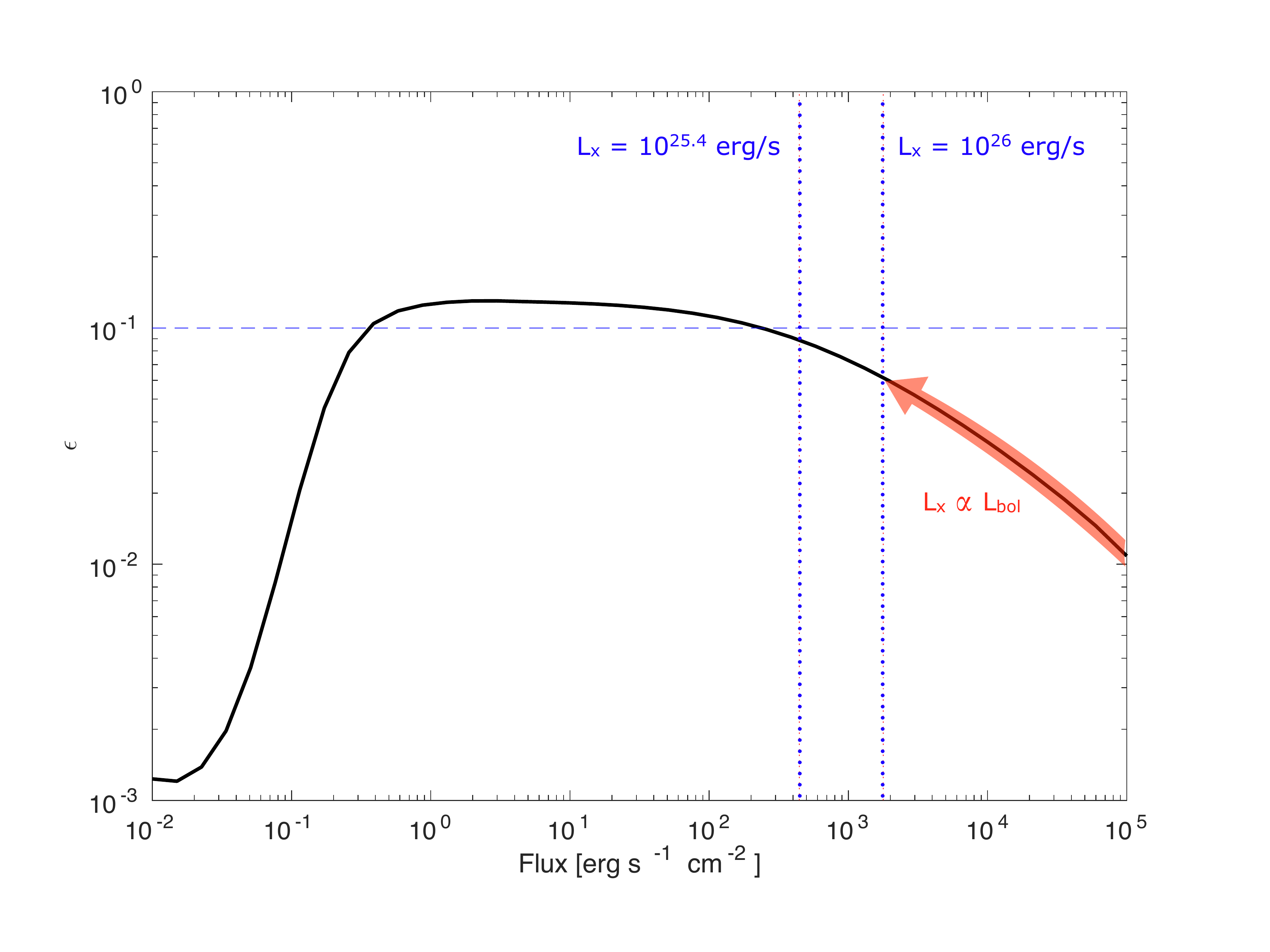}
	\caption{Variation of the efficiency $\epsilon$ with respect to the incoming flux obtained with the model from \citet{OwenAlvarez2016}, we note the stellar mass makes very little difference in the obtained mass-loss rates. The horizontal dashed line represents the value of $\epsilon$ usually used in energy-limited escape calculations. The vertical dotted line represents the XUV flux corresponding to $L_{\rm X} = 10^{25.4}$~erg/s for an orbital distance of $0.01$~au.}
	\label{efficacity_vs_flux}
	\end{center}
	\end{figure}

We can use the simulations to calculate the efficiency ($\epsilon$) parameter along with the corresponding mass loss.
Figure \ref{efficacity_vs_flux} shows the variation of $\epsilon$ with the incoming XUV flux, we note that the tidal gravity places a negligible role in changing the mass-loss rates with stellar mass.
At high fluxes the efficiency drops off due to the increased radiative cooling that can occur as the flows get more vigorous and dense. 
The drop off at low fluxes is simply caused by the fact that the heating rate is not strong enough to launch a powerful wind. 
At high fluxes the temperature peaks at a radii $\sim 1.2-1.4~\Rearth$, indicating some of the XUV photons are absorbed far from the planet, increasing its effective absorbing area, but at low fluxes the temperature peaks close to $1~\Rearth$.
Figure \ref{efficacity_vs_flux} shows that for a X-ray luminosity of $L_{\rm X} = 10^{25.4}$~erg/s (corresponding to a flux in XUV of $\sim450$~erg/s/cm$^2$), assuming an efficiency $\epsilon$ of 0.1 as is typically the case in energy-limited escape calculations overestimate the mass loss by a factor of $\sim1.17$.
	
Using these 1D radiation-hydrodynamic mass-loss simulations \citep[see Section \ref{hydrocode} and][]{OwenAlvarez2016}, we find that the temperature $T$ of the wind is of the order of $3000$~K, which is much lower than what is calculated for hot Jupiters \citep[of the order of $10^4$~K, e.g.,][]{Lammer2003, Erkaev2007, MurrayClay2009}.


\subsection{The joint escape of hydrogen and oxygen}\label{result_stoichio}

The computed mass loss $\dot{m}$ is linked to a mass flux $F_M$ given by: $F_M = \dot{m}/(4\pi \Rp^2)$.
We consider here a mass loss, but do not compute the proportion of hydrogen and oxygen atoms lost. 
Losing just hydrogen atoms and losing hydrogen and oxygen atoms does not have the same consequence for water loss. 
For example, if only hydrogen atoms are lost, losing an ocean means that the planet loses the mass of hydrogen contained in one Earth Ocean (9 times lower that the mass of water in one Earth Ocean).
This would change the proportion of H/O thus preventing any later recombination of water \citep[e.g. for Venus,][]{Gillmann2009}.
In contrast, if the planet loses hydrogen and oxygen in stoichiometric proportion, losing an ocean means that the planet is losing the mass of water contained in one Earth Ocean.
This is the more favorable case for water retention because it requires a higher energy to lose one ocean.

In the following, we estimate the proportion of escaping hydrogen and oxygen atoms.
The mass loss flux $F_M$ (kg.s$^{-1}$.m$^{-2}$) can be expressed in terms of the particle fluxes (atoms.s$^{-1}$.m$^{-2}$): 
\begin{equation}
F_M =  m_\textrm{H} F_\textrm{H} + m_\textrm{O} F_\textrm{O}.
\end{equation}

The ratio of the escape fluxes of hydrogen and oxygen in such an hydrodynamic outflow can be calculated following \citet{Hunten1987}:
\begin{equation}
r_\textrm{F}=\frac{F_\textrm{O}}{F_\textrm{H}} = \frac{X_\textrm{O}}{X_\textrm{H}} \frac{m_\textrm{c}-m_\textrm{O}}{m_\textrm{c}-m_\textrm{H}}, \label{equ:mc}
\end{equation}
where $m_\textrm{H}$ the mass of one hydrogen atom, $m_\textrm{O}$ the mass of one oxygen atom and $m_\textrm{c}$ is called the crossover mass and is defined by:
\begin{equation}\label{escape2}
m_\textrm{c} = m_\textrm{H} + \frac{kT F_\textrm{H}}{bg X_\textrm{H}},
\end{equation}
where $T$ is the temperature in the exosphere, $g$ is the gravity and $b$ is a collision parameter between oxygen and hydrogen.
In the oxygen and hydrogen mixture, we consider $X_\textrm{O}=1/3$, $X_\textrm{H}=2/3$, which corresponds to the proportion of dissociated water. 
This leads to $X_\textrm{O}/X_\textrm{H}=1/2$. 

When $m_\textrm{c} < m_\textrm{O}$, only hydrogen atoms are escaping and $F_\textrm{H} = F_M/m_\textrm{H}$.
When $m_\textrm{c} > m_\textrm{O}$, hydrogen atoms drag along some oxygen atoms and:
\begin{equation}\label{escape3}
F_M = (m_\textrm{H} + m_\textrm{O}r_\textrm{F})F_\textrm{H}.
\end{equation}

With Equations \ref{equ:mc}, \ref{escape2} and \ref{escape3}, we can compute the mass flux of hydrogen atoms:
\begin{eqnarray}\label{escape4}
F_\mathrm{H} &=&\frac{F_M  +  m_\mathrm{O}X_\mathrm{O}\left(m_\mathrm{O}-m_\mathrm{H}\right)\frac{b g}{k T } 
}{m_\mathrm{H}  + m_\mathrm{O}\frac{X_\mathrm{O}}{X_\mathrm{H}}}.
\end{eqnarray}

In order to calculate the flux of hydrogen atoms, we need an estimation of the XUV luminosity of the star considered, as well as an estimation of the temperature $T$.
We use the estimations of the XUV flux of Section \ref{phys_input} and the estimation of $\epsilon$ and $T$ obtained with the 1D radiation-hydrodynamic mass-loss simulations \citep{OwenAlvarez2016}.

We assume that the XUV luminosity is here $L_0 = 5 \times L_{\rm X} = 5\times10^{25.4}$~erg/s, that the planet is at $a =0.013~$au and following Section \ref{hydrocode} (Figure \ref{efficacity_vs_flux}), we assume here $\epsilon = 0.097$. 
This is our baseline case.
Using these values and Equation \ref{eqmassloss1}, we calculate the mass loss flux $F_M = F_0$ for a $1~\Mearth$ planet:
\begin{equation}
\label{F0}
F_0 = \frac{\dot{m}}{4\pi \Rp^2} = \epsilon \frac{F_{{\rm XUV}}\Rp}{4 \G \Mp (a/1 {\rm au})^2} = 1.02 \times 10^{-10}~\mathrm{kg.s}^{-1}\mathrm{m}^{-2}.
\end{equation}

\begin{figure}
	\begin{center}
	\includegraphics[width=\linewidth]{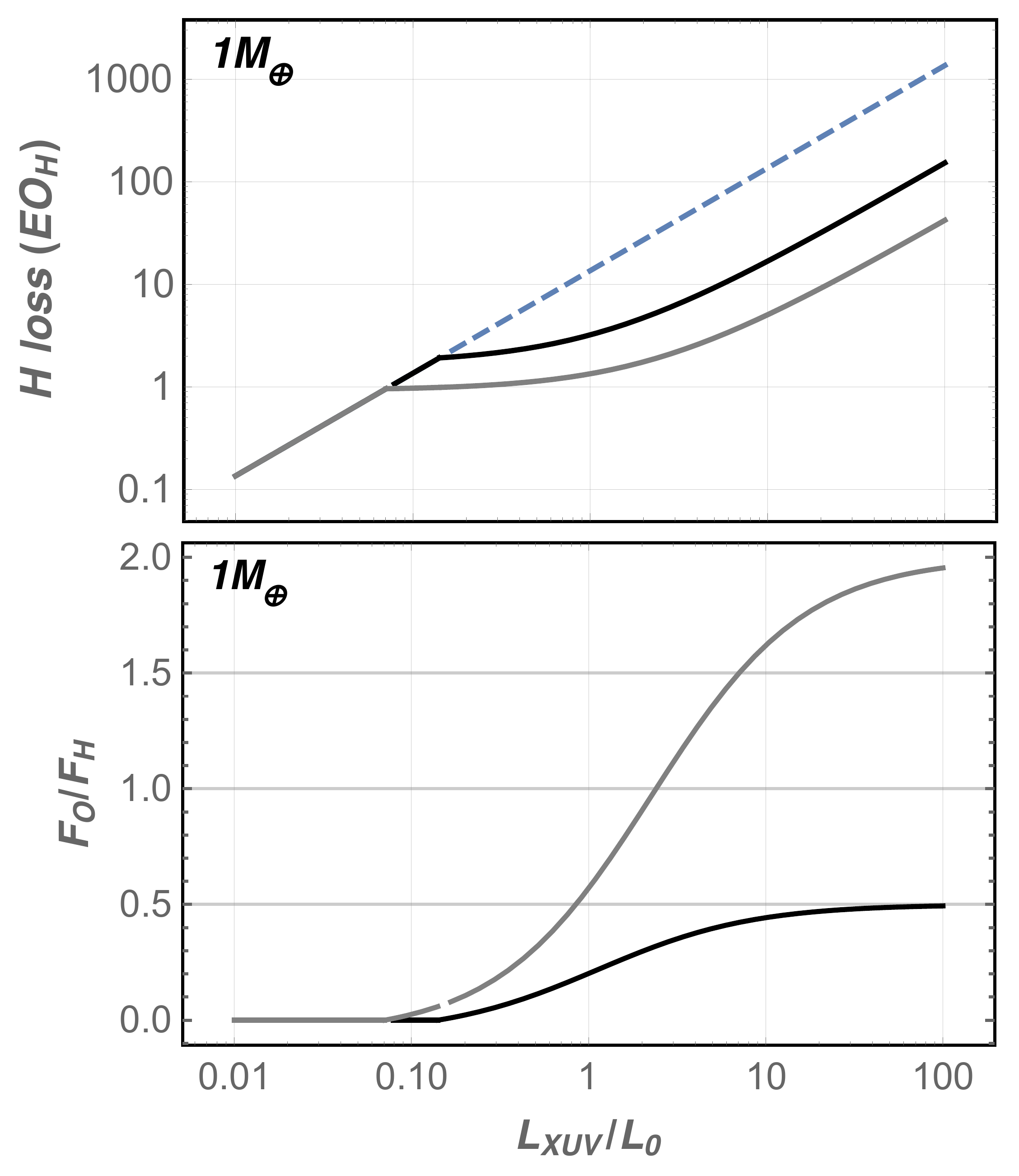}
	\caption{Top panel: Mass of hydrogen lost during the runaway phase as a function of the XUV luminosity for an Earth-like planet at $0.013~$au of a $0.08~\Msun$ dwarf. The mass loss are in unit of Earth Ocean equivalent content of hydrogen ($EO_H$). The blue dashed line is for the limit case where only hydrogen atoms escape, which is valid at low irradiation, when the escape is not strong enough to drag oxygen atoms. The black line accounts for oxygen dragging in the case where $X_\mathrm{O}=X_\mathrm{H}/2=1/3$. The gray line is the same for an atmosphere slightly depleted in hydrogen ($X_\mathrm{O}=2X_\mathrm{H}=2/3$). Bottom panel: Ratio of the flux of oxygen atoms over the flux of hydrogen atoms. The colors are the same as for the top panel. $r_\textrm{F} = F_\mathrm{O}/F_\mathrm{H} = 0.5$ means that there is stoichiometric loss of hydrogen and oxygen. For the baseline case, $r_\textrm{F} = 0.20$.}
	\label{FH_vs_FtotonF0}
	\end{center}
	\end{figure}

Using Equation \ref{escape4} and the estimation of $T$, we can compute the flux of hydrogen atoms $F_{\rm H}$ as a function of the XUV luminosity $L_{\rm XUV}$ compared to our baseline case luminosity $L_0$.
Figure \ref{FH_vs_FtotonF0} shows the behavior of the mass loss of the atmosphere in units of Earth Ocean equivalent content of hydrogen ($EO_H$) with respect to the ratio $L_{\rm XUV}/L_0$.
This behavior is plotted for two cases: a stoichiometric mixture of hydrogen and oxygen atoms ($X_\mathrm{O}=X_\mathrm{H}/2=1/3$) and a mixture slightly depleted in hydrogen ($X_\mathrm{O}=2X_\mathrm{H}=2/3$).
We can see that the mass loss is a monotonous function of the total XUV luminosity, and for both cases it saturates much below our baseline case.
For a stoichiometric mixture, at a ratio $L_{\rm XUV}/L_0 \sim 0.13$, the oxygen atoms start to be dragged along therefore consuming energy.
For a mixture depleted in hydrogen, oxygen atoms start to be dragged along for lower incoming XUV luminosity $L_{\rm XUV}/L_0 \sim 0.075$.

Here however, for $L_{\rm XUV}=L_0$, we find that one oxygen atom is lost for about 5 hydrogen atoms: $r_\textrm{F} = F_\mathrm{O}/F_\mathrm{H} = 0.20$.
To conclude, for a X-ray luminosity of $10^{25.4}$~erg/s, there is no stoichiometric loss of hydrogen and oxygen. 
However, the situation is not as catastrophic as it would be if only hydrogen escaped. 
In the following, we give the hydrogen lost by the planet.
We assume that $r_\textrm{F}$ is constant with time for a given value of the XUV flux, which is equivalent to assume an infinite reservoir of water. 
With a finite water reservoir one has to account for the evolution of the O/H ratio. 
As it makes the result depend on the initial reservoir, we choose to maximize the effective water loss by using the value of $r_\textrm{F}$ calculated for a ratio $X_\mathrm{O}=X_\mathrm{H}/2$.
The hydrogen mass loss is thus given by:
\begin{equation}
\label{Hyd_loss}
M_{\rm H} = m \frac{m_\textrm{H}}{m_\textrm{H} + r_\textrm{F} m_\textrm{O}},
\end{equation}
where $m$ is obtained with Equation \ref{eqmassloss2} and is given in units of the mass of hydrogen in one Earth Ocean ($EO_H$, which corresponds to $\sim 1.455\times10^{20}$~kg).
Hydrogen is the limiting element for the recombination of water, so that the remaining content of hydrogen in $EO_H$ actually represents the ocean portion available for precipitation once in the HZ.

According to Equation \ref{escape2}, $r_\textrm{F}$ depends on the gravity of the planet considered. 
In the following, we also calculate the hydrogen loss from planets of $0.1~\Mearth$ and $5~\Mearth$.
Computing the hydrogen loss for these cases show us that the higher the mass the lower is $r_\textrm{F}$.
For $L_{\rm XUV}=L_0$, we find that $r_\textrm{F} \sim 0.45$ for the $0.1~\Mearth$ planet (meaning that the loss is quasi-stoichiometric), and $r_\textrm{F} = 0$ for the $5~\Mearth$ planet (meaning that the only hydrogen atoms escape).


\section{Water loss of planets in the UCD habitable zone}\label{ener_lim_esc}

Using the estimation of $\epsilon$ as a function of XUV flux (Figure \ref{efficacity_vs_flux}), the estimations of $r_\textrm{F}$ as calculated in the previous section and the equations of Section \ref{method}, we can now estimate the loss of hydrogen from planets around UCDs.
	
\begin{figure}
	\begin{center}
	\includegraphics[width=\linewidth]{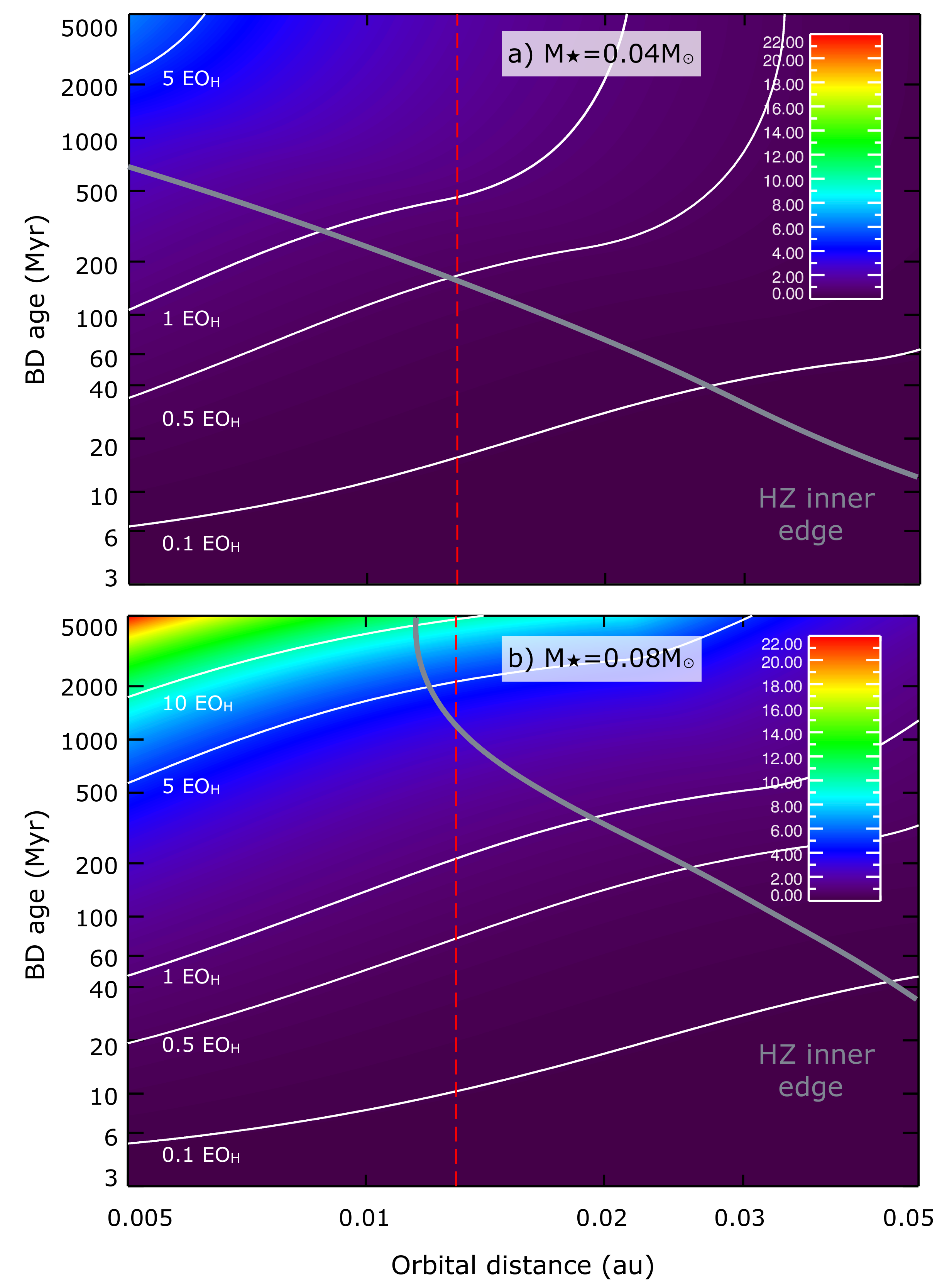}
	\caption{Map of hydrogen loss (in $EO_H$) as a function of the age of the BD of a) $0.04~\Msun$ and b) $0.08~\Msun$ and the orbital distance of the planet. White lines correspond, from left to right, to a loss of $10~EO_H$, $5~EO_H$, $1~EO_H$, $0.5~EO_H$ and $0.1~EO_H$. The grey line corresponds the inner edge of the HZ: the HZ lies above this line. Here the X-ray luminosity evolves as $10^{-5}\times L_{{\rm bol}}$.}
	\label{maps_massloss_Mp1_Ms_alum5}
	\end{center}
	\end{figure}

Figure \ref{maps_massloss_Mp1_Ms_alum5}a) shows the evolution of the hydrogen loss from an Earth-mass planet orbiting a $0.04~\Msun$ BD as a function of both time and orbital radius (assumed to remain constant). 
We assume that the X-ray luminosity scales as $10^{-5}~L_{{\rm bol}}$ and we calculate $\epsilon$ and $r_\textrm{F}$ accordingly (typically, for a planet at 0.013~au, $\epsilon$ varies between      0.045 and 0.111 as the XUV flux decreases and $r_\textrm{F}$ decreases from 0.43 to 0.10).
Let us consider that the planet is located at $0.013$~au from a $3~$Myr-old BD.
Figure \ref{maps_massloss_Mp1_Ms_alum5}a) shows that it would lose only $0.48~EO_H$ in the 156~Myr before entering the HZ (vertical red dashed line). 
This planet could therefore be considered a potentially habitable planet, assuming an appropriate atmospheric composition and structure.
As the bolometric luminosity decreases with time, the XUV luminosity decreases as well, which has the effect of increasing the efficiency $\epsilon$ (Figure \ref{efficacity_vs_flux}) and decreasing $r_\textrm{F}$.
Consequently, with this compensation, the efficiency of the mass loss do not decrease linearly with decreasing XUV-flux. 
Figure \ref{maps_massloss_Mp1_Ms_alum5}b) shows the evolution of the hydrogen loss from an Earth-mass planet orbiting a $0.08~\Msun$ BD.
A planet located at $0.013$~au from a $3~$Myr-old BD lose $\sim 3.2~EO_H$ in the 1180~Myr before entering the HZ (vertical red dashed line). 

\begin{figure*}[b]
	\centering
	\includegraphics[width=\linewidth]{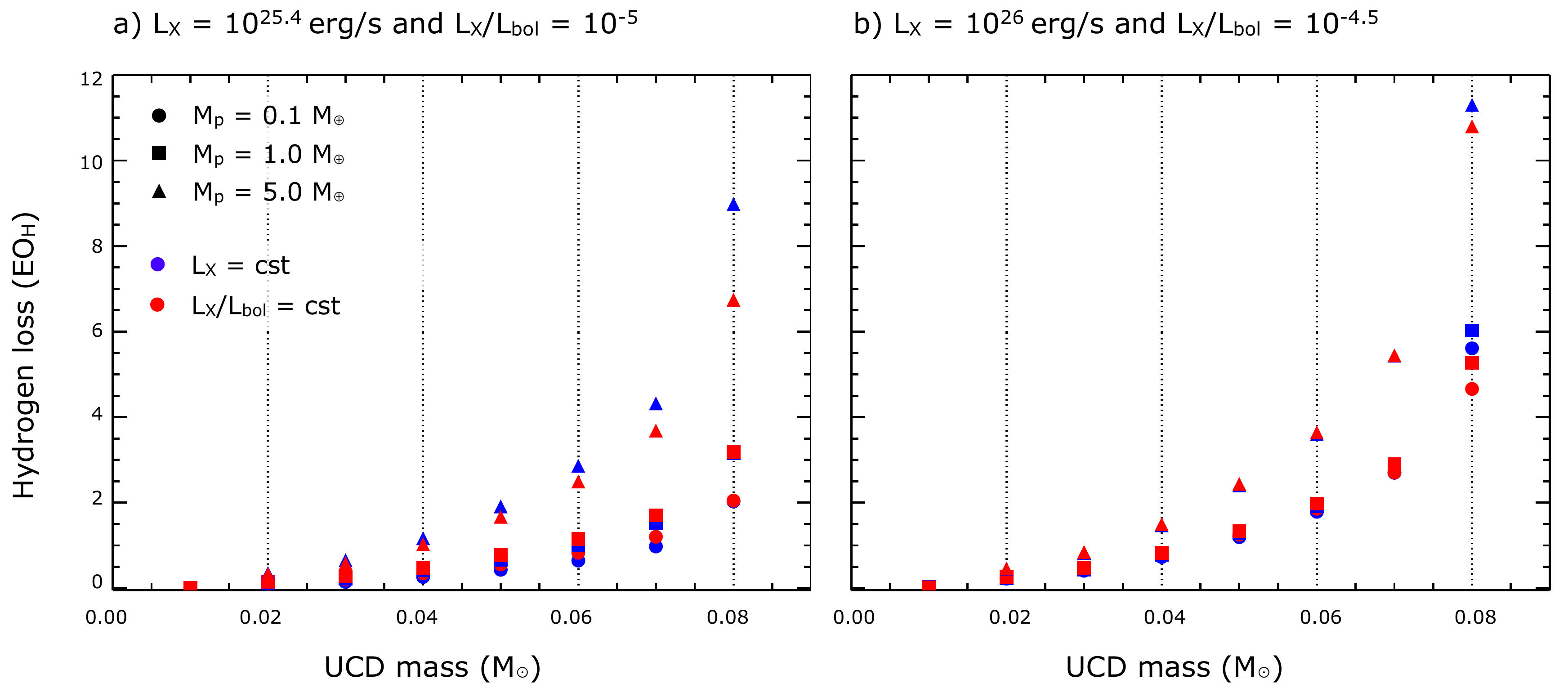}
	\caption{Hydrogen loss (in units of Earth's ocean) as a function of the mass of the UCD for a planet of different masses at $0.013$~au. Panel a) corresponds to $L_{\rm X} = 10^{25.4}$~erg/s (in blue) and $L_{\rm X}/L_{{\rm bol}}=10^{-5}$ (in red). Panel b) corresponds to $L_{\rm X} = 10^{26}$~erg/s (in blue) and $L_{\rm X}/L_{{\rm bol}}=10^{-4.5}$ (in red). }
	\label{massloss_Mp_alum5cst_Mbd_legend}
	\end{figure*}

Figure \ref{massloss_Mp_alum5cst_Mbd_legend}a) shows the hydrogen loss for planets of $0.1~\Mearth$, $1~\Mearth$ and $5~\Mearth$ orbiting UCDs of different masses at $0.013$~au. 
The figure shows how much water was lost by the time each planet reached the HZ. 
This time is $\sim 8$~Myr for a BD of $0.01~\Msun$ and $\sim 1180$~Myr for a dwarf of $0.08~\Msun$.
The calculations were done for both cases: $L_{\rm X}/L_{{\rm bol}}=10^{-5}$ (red symbols) and $L_{\rm X} = 10^{25.4}$~erg/s (blue symbols).
As before, $\epsilon$ and $r_\textrm{F}$ were calculated consistently for each planet and for the two X-ray luminosity assumptions.

	\begin{figure*}
	\centering
	\includegraphics[width=\linewidth]{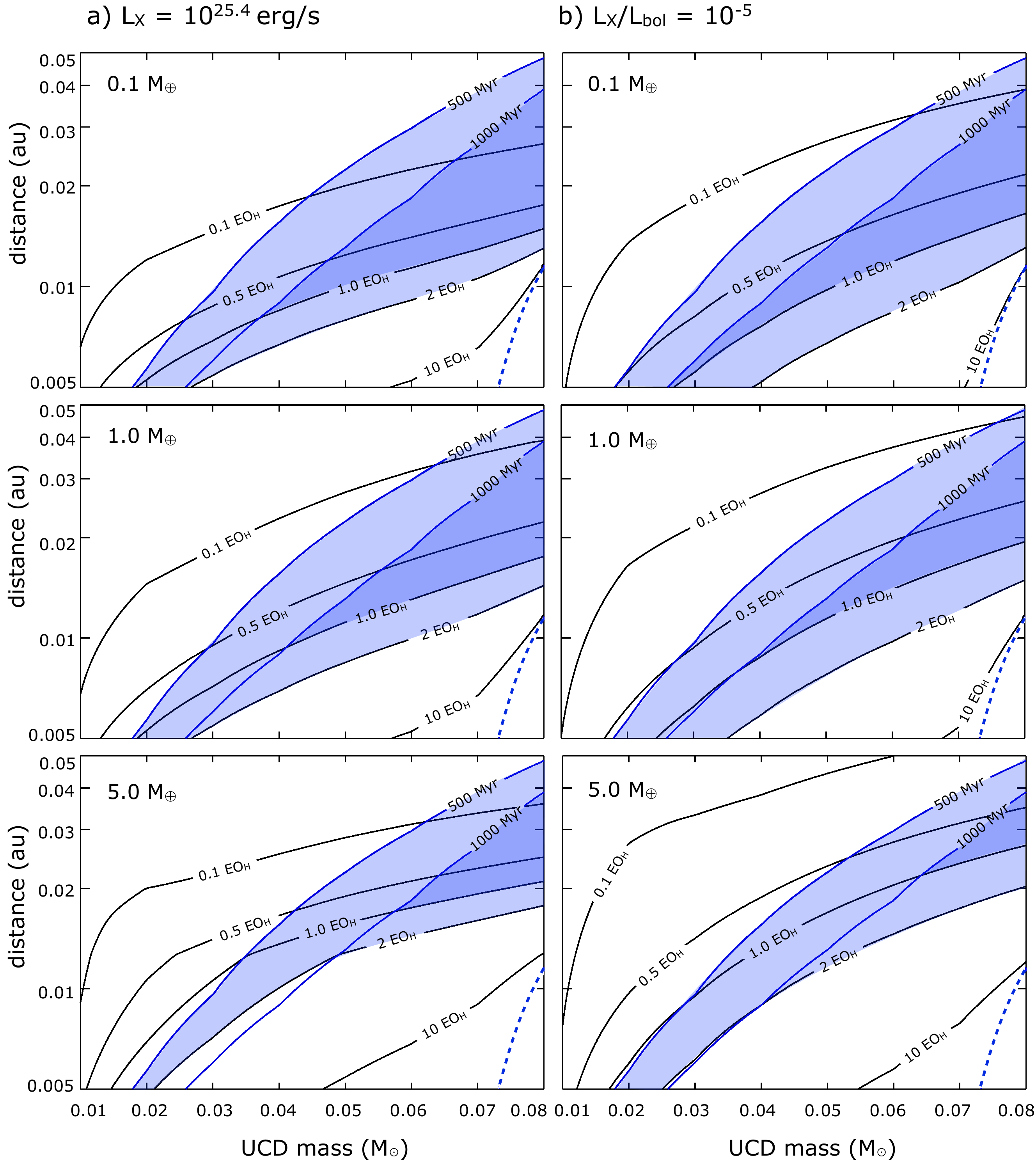}
	\caption{Hydrogen loss (black contour lines) as a function of the mass of the UCD and the orbital distance of the planet. Panels a) correspond to $L_{\rm X} = 10^{25.4}$~erg/s and panels b) correspond to $L_{\rm X}/L_{{\rm bol}}=10^{-5}$. The blue lines correspond from left to right to a time spent in the HZ of 500~Myr and 1~Gyr. The dashed blue line corresponds to the limit where the planets never reach the HZ because the dwarf initiated the fusions of hydrogen preventing the inner edge of the HZ to sweep in towards very small orbital distances (see Figure \ref{aM004_Mp1_eo001}). The blue shaded areas represent two interesting parameter spaces: the planets in the light blue area lose less than $2~EO_H$ before reaching the HZ and they will spend more than 500~Myr in it, the planets in the dark blue area lose less than $1~EO_H$ before reaching the HZ and they will spend more than 1~Gyr in it.}
	\label{maps_massloss_tinit_3000000_X5_EOH_rf_HZ_epsOwen_normal_1}
	\end{figure*}

We find that planets orbiting more massive UCDs lose more water. 
This is mainly because of the much longer time spent by those planets interior to the HZ. 
Planets at 0.013~au around BDs of mass $0.01~\Msun$ lose less than 0.04~$EO_H$, whatever their mass. 
Planets at 0.013~au around BDs of mass $0.05~\Msun$ lose less than 2~$EO_H$, whatever their mass.
Planets at 0.013~au around UCDs of mass $0.08~\Msun$ lose more than 2~$EO_H$, whatever their mass.

We find that higher-mass planets lose their water at a higher rate than lower-mass ones. 
This is due to the effect of gravity on the cross-over mass that makes $r_\textrm{F}$ smaller for a higher gravity. 
In other words too high a gravity prevents the loss of oxygen and thus enhances the loss of hydrogen (which is not directly affected by gravity, e.g, \citealt{LugerBarnes2015}).
Consequently, throughout the evolution, we expect more massive planets to lose more than low-mass planets.
For example, assuming $L_{\rm X}/L_{{\rm bol}}=10^{-5}$, we find that the $5~\Mearth$ planet orbiting a UCD of $0.08~\Msun$ loses 6.7~$EO_H$ before reaching the HZ, while the $1~\Mearth$ loses only 3.2~$EO_H$ and the $0.1~\Mearth$ only 2.0~$EO_H$.

Assuming $L_{\rm X}/L_{{\rm bol}}=10^{-5}$, we find that $1~\Mearth$ planets orbiting at 0.013~au around BDs with masses smaller than $0.06~\Msun$ lose less than 1~$EO_H$ before reaching the HZ and $1~\Mearth$ planets orbiting BDs of mass $\lesssim 0.07~\Msun$ lose less than 2~$EO_H$ before reaching the HZ.
Whatever the mass of the UCD, low mass planets ($\Mp \leq 1~\Mearth$) lose less than 3.2~$EO_H$ before reaching the HZ.
Whatever the mass of the UCD, whatever the mass of the planet and the XUV-luminosity assumption, all planets lose less than 9~$EO_H$ before reaching the HZ.

Figure \ref{maps_massloss_tinit_3000000_X5_EOH_rf_HZ_epsOwen_normal_1} shows the hydrogen loss as a function of the planet's orbital distance and mass of host UCD. 
The black lines represent different hydrogen loss levels ($1, 2, 10~EOH_H$) and the white lines represent levels of time the planet passes in the HZ (as computed as in Figure \ref{aM004_Mp1_eo001}).
The closer the planet and the more massive the UCD, the more hydrogen is lost.
The closer the planet and the more massive the UCD, the more time the planet spends in the HZ.
However for the more massive UCD we consider (as can be seen in Figure \ref{aM004_Mp1_eo001} for the UCD of $0.08~\Msun$), the planets on the closest orbits are always interior to the HZ and thus stay in a runaway phase during all the time of the evolution (10~Gyr).
In Figure \ref{maps_massloss_tinit_3000000_X5_EOH_rf_HZ_epsOwen_normal_1}, these planets are separated from the rest by the blue dashed line.
They can lose up to $160~EO_H$.
There is a compromise to be found between the hydrogen loss and the time the planet spends in the HZ: planets around low mass BDs lose little hydrogen but they stay a short time in the HZ, planets around the higher mass UCDs considered here spend a longer time in the HZ but they lose more hydrogen prior to entering.
The shaded regions in Figure \ref{maps_massloss_tinit_3000000_X5_EOH_rf_HZ_epsOwen_normal_1} shows an interesting parameter space for each planet and XUV emission hypothesis: the planets in these regions lose less than $1~EO_H$ before entering the HZ and spend in the HZ more than 1~Gyr.
Planets with a similar or larger water content as the Earth would thus enter the HZ with enough water to form oceans and as they spend a long time in the HZ, this gives time for life to eventually appear and modify the environment \citep[][]{Bolmont2011}
For example, for an Earth mass planet and assuming $L_{\rm X}/L_{{\rm bol}}=10^{-5}$, we find that the planets around UCDs more massive than $\sim0.035~\Msun$ and farther away than $0.007$~au fulfill these conditions. 
Of course, when considering higher mass UCDs, the minimum orbital distance increases: a planet around a $0.08~\Msun$ UCD has to be between farther away than $\sim0.02$~au to fulfill the conditions.
If we consider softer constraints, for example cases for which the planets lose less than $2~EO_H$ and spend more than 500~Myr in the HZ, the parameter space gets much bigger: for example, Earth-mass planets as close as $0.005$~au around a $0.02~\Msun$ BD fulfill these conditions.
However these very close-in planets could be in danger of falling onto the BD: they could be interior to the corotation radius \citep[see Figure \ref{aM004_Mp1_eo001} and][]{Bolmont2011}.
If we consider $5~\Mearth$ planets, the parameter space corresponding to a loss $\leq 1~EO_H$ and a time in the HZ $\geq 1$~Gyr shrinks towards the higher UCD masses and bigger orbital distances.
If we consider $0.1~\Mearth$ planets, the parameter space corresponding to a loss $\leq 1~EO_H$ and a time in the HZ $\geq 1$~Gyr extends towards the lower UCD masses and smaller orbital distances.
We thus can conclude that with favorable atmospheric conditions and reasonable water content (a few oceans), Earth-mass planets between $0.01$ and $0.04$~au orbiting UCDs of mass $0.04$~--~$0.08~\Msun$ would be good targets for the characterization of a potentially habitable planet.

\medskip

These calculations were made with the following assumptions on the X-ray luminosity of these faint dwarfs: $L_{\rm X}/L_{{\rm bol}}=10^{-5}$ and $L_{\rm X} = 10^{25.4}$~erg/s.
These values are probably a good approximation of the X-ray flux received by planets around BDs \citep{Berger2010, CookWilliamsBerger2014, Williams2014}. 
However, one might want to consider, for the higher mass UCDs we consider here, stronger high-energy irradiation levels, such as $L_{\rm X}/L_{{\rm bol}}=10^{-4.5}$ and $L_{\rm X} = 10^{26}$~erg/s, which correspond to the quiescent emission of the object 2MASS~14542923+1606039 Bab \citep[see][]{Williams2014}.
Figure \ref{massloss_Mp_alum5cst_Mbd_legend}b) shows the hydrogen loss for these two X-ray luminosity cases. 
The hydrogen loss is higher than before.
Due to the dependance of the efficiency $\epsilon$ and of $\rf$ with the XUV luminosity, the mass loss computed with $L_{\rm X}/L_{{\rm bol}}=10^{-4.5}$ is less than $\sim 3.16$ higher than the one calculated with $L_{\rm X}/L_{{\rm bol}}=10^{-5}$.
Similarly, the loss computed with $L_{\rm X} = 10^{26}$~erg/s is less than $\sim 4$ times higher than the one computed with $L_{\rm X} = 10^{25.4}$~erg/s.
Using these values, we find that all planets orbiting BDs of mass lower than $0.03~\Msun$ at 0.013~au lose less than 1~$EO_H$ before reaching the HZ.
Note that when the XUV flux is high (red symbols), $r_\textrm{F}$ for the two lower mass planets tend to 0.5, counterbalancing the effect of the planet's gravity on the crossover mass and leading here to a very similar mass loss.

With these higher X-ray luminosity assumptions, we thus can conclude that with favorable atmospheric conditions and reasonable water content (a few oceans), Earth-mass planets between $0.018$ and $0.04$~au (instead of $0.01$~--~$0.04$~au) orbiting UCDs of mass $0.06$~--~$0.08~\Msun$ (instead of $0.04$~--~$0.08~\Msun$) would be good targets for the characterization of a potentially habitable planet.

\section{Implication for the TRAPPIST-1 planets}

The three planets of the TRAPPIST-1 system \citep{Gillon2016} are Earth-sized planets, and thus probably rocky \citep{WeissMarcy2014, Rogers2015}. 
They orbit a M8-type dwarf of $0.080 \pm 0.009~\Msun$\footnote{TRAPPIST-1 is therefore not a BD but a very low mass star.}
TRAPPIST-1b is located at $a_b = 0.011$~au, TRAPPIST-1c at $a_c = 0.015$~au. 
The orbit of TRAPPIST-1d is poorly constrained, however it is farther away than $a_d = 0.022$~au.
The irradiation of the planets are respectively: 4.25~$S_\oplus$, 2.26~$S_\oplus$ and $0.02$--$1~S_\oplus$, where $S_\oplus$ is the insolation received by the Earth. 
Therefore TRAPPIST-1d could be in the HZ.
The age of the system has been estimated to be more than $500$~Myr.
The structural evolution grids we use in this article for a dwarf star of $0.08~\Msun$ \citep{ChabrierBaraffe1997} show that the luminosity and radius of TRAPPIST-1 correspond to a body of $\sim 400~$Myr, which is lower than the estimated age of the system. 
This is consistent with the fact that evolution models seem to under-estimate the luminosity of low-mass objects \citep{Chabrier2007}.
However, we explored the mass range allowed by the observations and found that a dwarf star of $0.089~\Msun$ can reproduce the characteristics of \Tpp~at an age of $\sim850$~Myr.

\begin{figure*}
	\centering
	\includegraphics[width=17cm]{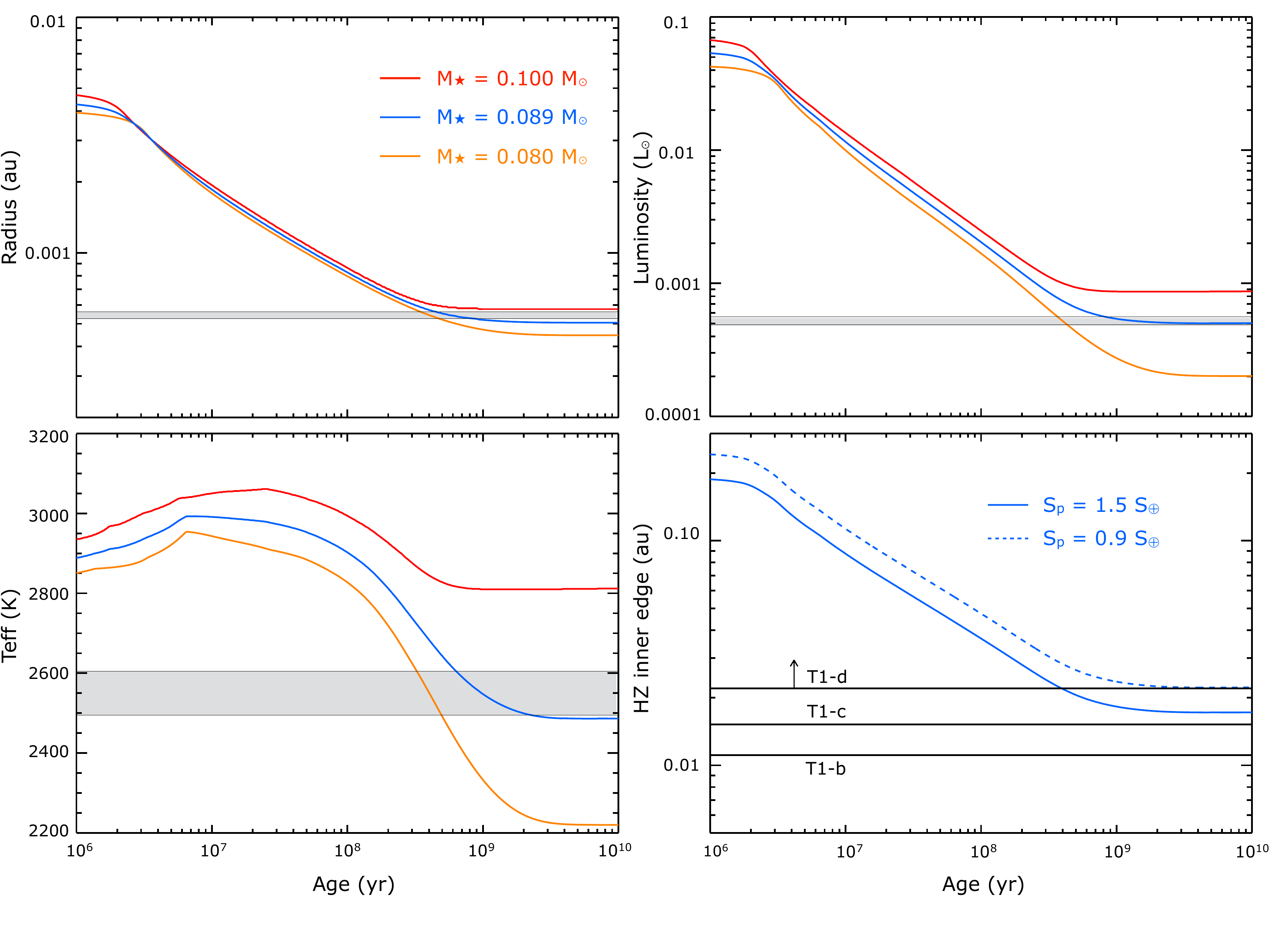}
	\caption{Characteristics of the star's model we used for the calculation of hydrogen loss. Top left: Evolution of the radius for the 2 models \citep[in red and orange][]{ChabrierBaraffe1997} and our interpolation (in blue). The grey area corresponds to $\Rs = 0.117\pm0.004~\Rsun$ \citep{Gillon2016}. Top right: Evolution of the luminosity, the grey area corresponds to $\Ls = 0.000525\pm0.000036~\Lsun$. Bottom left: Evolution of the effective temperature, the grey area corresponds to $T_{\rm eff} = 2550\pm55~$K. Bottom right: Evolution of the inner edge of the HZ for two different assumptions: $S_p = 1.5~\Searth$ and $S_p = 0.9~\Searth$. The orbital distances of the three planets of \Tpp~are also represented.}
	\label{charac_trappist_model}
	\end{figure*}

Figure \ref{charac_trappist_model} shows the evolutionary tracks we used to simulate the luminosity evolution of \Tpp. 
We interpolated the values of radius, luminosity and effective temperature between the evolutionary tracks of a $0.08~\Msun$ dwarf and a $0.1~\Msun$ dwarf \citep{ChabrierBaraffe1997}.
Figure \ref{charac_trappist_model} shows that the characteristics of the star -- radius, luminosity and effective temperature -- can be reproduced with our interpolated tracks for ages between 800~Myr and 900~Myr, which is compatible with the estimation of the age of the star made by \citet{Gillon2016}.

We use here two different assumptions to calculate the HZ inner edge: $S_p = 0.9~\Searth$, which corresponds to the inner edge for a non-synchronized planet \citep{Kopparapu2013} and $S_p = 1.5~\Searth$ which corresponds to the inner edge for a synchronized planet \citep{Yang2013}.
Following this model, the two inner planets of \Tpp~always stay interior to the HZ.
As the orbit of planet d is poorly constrained, we considered three different orbits: 0.022~au (the closest one), 0.058~au (the most probable one) and 0.146~au (the farthest one).
A planet at $0.022$~au enters the HZ corresponding to $S_p = 1.5~\Searth$ at an age of $393$~Myr (later called $T_{\rm HZ} (1.5~\Searth)$).
However, it never enters the HZ corresponding to $S_p = 0.9~\Searth$.
For a planet at $0.058$~au, $T_{\rm HZ} (1.5~\Searth) = 29$~Myr and $T_{\rm HZ} (0.9~\Searth) = 58$~Myr.
For a planet at $0.146$~au, $T_{\rm HZ} (1.5~\Searth) = 3.2$~Myr and $T_{\rm HZ} (0.9~\Searth) = 5.4$~Myr.
If we had considered more massive stars than what is allowed by \citet{Gillon2016}, the entry in the HZ would have been postponed by at least a few tens of million years increasing the period of time the planet spends in the runaway phase.

\medskip
To calculate the mass loss from the planets of the TRAPPIST-1 system, we assumed different XUV emissions.
As explained in Section \ref{phys_input}, TRAPPIST-1 is part of the transition population between early M type and late M, early L.
In order to treat the whole XUV range possible we assumed the two different XUV luminosity measured by \citet{Wheatley2016}: 
\begin{description}
\item[-] $L_{\rm X}/L_{{\rm bol}}=10^{-3.7}$ and $L_{\rm X}/L_{{\rm bol}}=10^{-3.4}$\\
Observational studies \citep[e.g.,][]{WilliamsCookBerger2014, CookWilliamsBerger2014} indicate a significantly large scatter at spectral type M8, with values ranging between $L_{\rm X}/L_{\rm bol} =10^{-5}$ and $10^{-3}$ in quiescence. 
\citet{CookWilliamsBerger2014} mention another analog of \Tpp~, LP~412-31\footnote{Its bolometric luminosity $L_{\rm bol}/\Lsun \sim 10^{-3.29}$ is close to the luminosity of \Tpp~($L_{\rm bol}/\Lsun \sim 10^{-3.28}$), its $V \sin i$ of 8~km.s$^{-1}$\citep{Reid2002, Newton2016} is also close to \Tpp's $\sim 6$~km.s$^{-1}$.}, which has a quiescent emission of $10^{27.2}$~erg.s$^{-1}$ or $L_{\rm X}/L_{\rm bol} =10^{-3.1}$.
Furthermore, using XMM-Newton observations, \citet{Wheatley2016} measured recently for TRAPPIST-1 $L_{\rm X}/L_{\rm bol} =10^{-3.7}$ to $10^{-3.4}$, which is significantly higher than the value we adopted for the UCDs of the previous Sections.
\item[-] $L_{\rm X}/L_{{\rm bol}}=10^{-5}$\\
This is what we used for UCDs in the previous Sections.
As the measurements of \citet{Wheatley2016} could be due to a flare, we consider this much lower flux.
This also corresponds to an analog of TRAPPIST-1: the M8 dwarf VB~10\footnote{Its bolometric luminosity $L_{\rm bol}/\Lsun \sim 10^{-3.3}$ is close to the luminosity of \Tpp~($L_{\rm bol}/\Lsun \sim 10^{-3.28}$), its $V \sin i$ of 6.5~km.s$^{-1}$ is also close to \Tpp's $\sim 6$~km.s$^{-1}$. Their radii are also similar within a few percent.}.
Observations of VB~10 by \citet{Fleming2000} showed that the quiescent emission was $L_{\rm X} = 10^{-5.0}~L_{\rm bol}$ and the flaring emission was $L_{\rm X} = 10^{-2.8}~L_{\rm bol}$.
Later, \citet{Berger2008} measured $L_{\rm X} = 10^{-5.0}~L_{\rm bol}$ for the quiescent emission and $L_{\rm X} = 10^{-4.1}~L_{\rm bol}$ during flaring events.
Finally, \citet{WilliamsCookBerger2014} and \citet{CookWilliamsBerger2014} found that the quiescent emission of VB~10 was $L_{\rm X} = 10^{-5.1}~L_{\rm bol}$ and that the flaring emission was $L_{\rm X} = 10^{-4.6}~L_{\rm bol}$.\\
\end{description}

	\begin{figure*}
	\centering
	\includegraphics[width=\linewidth]{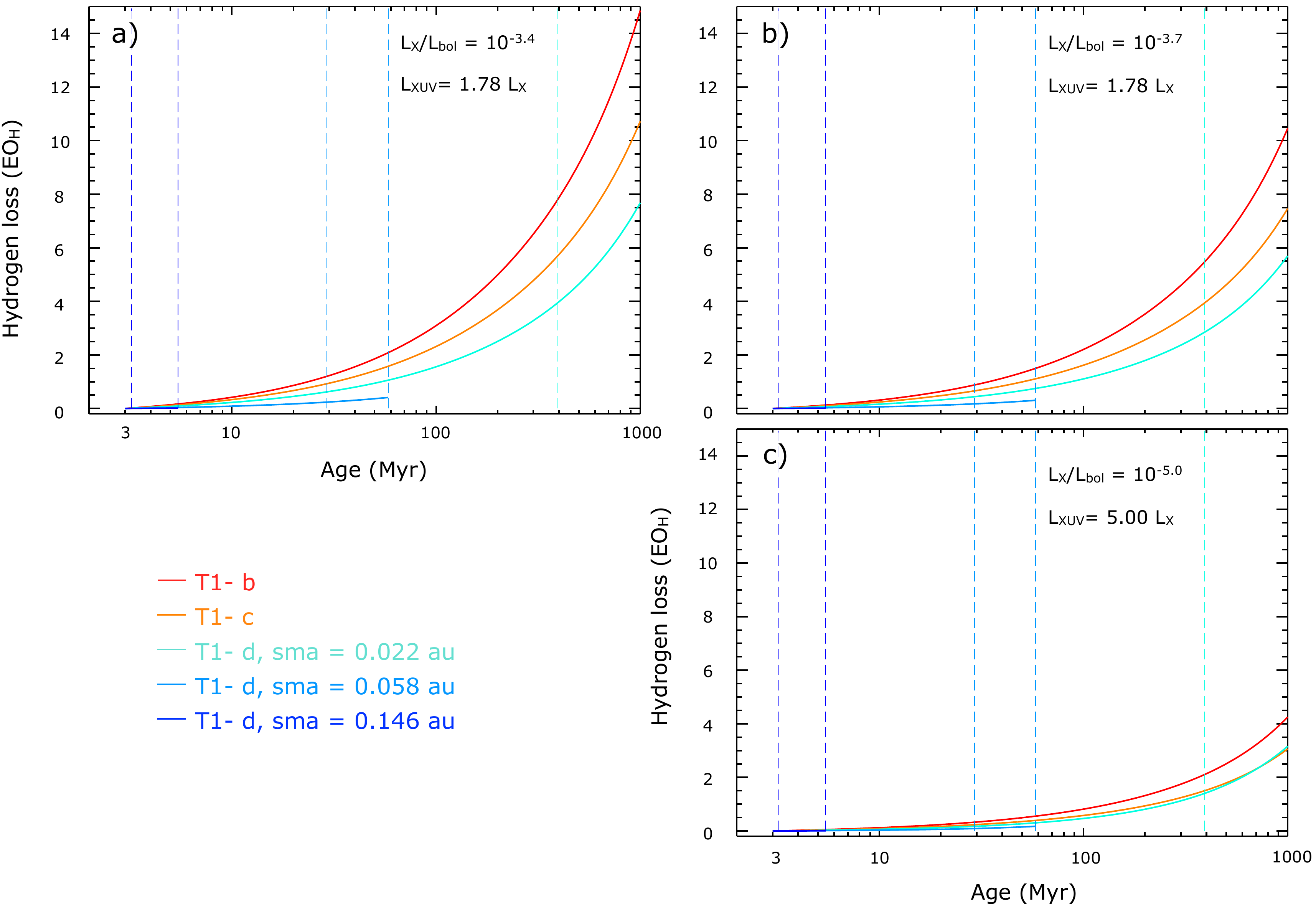}
	\caption{Hydrogen loss as a function of time for the planets of TRAPPIST-1. Panel a) corresponds to $L_{\rm X}/L_{{\rm bol}}=10^{-3.4}$, panel b) corresponds to $L_{\rm X}/L_{{\rm bol}}=10^{-3.7}$ and the panel c) to $L_{\rm X}/L_{{\rm bol}}=10^{-5.0}$. The results were obtained with Equation \ref{eqmassloss2} for the radiuses and semi-major axes of the planets given by \citet{Gillon2016}. The dashed vertical lines represents the time the planets reach the HZ for both assumptions.}
	\label{massloss_TRAPPIST_0089_bcddd_alum_tinit_3_EOH_newrf_lin}
	\end{figure*}

\begin{table*}
\centering
\caption{Lost hydrogen in $EO_H$ and corresponding O$_2$ pressure for the TRAPPIST-1 planets at the time they enter the HZ (with the two assumptions about the inner edge). The numbers in bold correspond to cases for which the planet never reaches the HZ, the hydrogen loss is then given for the age of the star ($\sim 850$~Myr, according to our model). The two values indicated correspond to $t_0 = 10$~Myr and $t_0 = 3$~Myr.}
\vspace{0.1cm}
\begin{tabular}{|c|c|c|c|c|c|c|c|c|c|c|}
\hline
& & \multicolumn{6}{|c|}{Hydrogen loss ($EO_H$)} & \multicolumn{2}{|c|}{$P_{\rm O_2}$(bar)}\\
& & \multicolumn{2}{|c|}{$L_{\rm X}/L_{\rm bol} =10^{-3.4}$} & \multicolumn{2}{|c|}{$L_{\rm X}/L_{\rm bol} =10^{-3.7}$} & \multicolumn{2}{|c|}{$L_{\rm X}/L_{\rm bol} =10^{-5.0}$}& \multicolumn{2}{|c|}{}\\
& SMA (au) & $T_{\rm HZ} (0.9~\Searth)$ & $T_{\rm HZ} (1.5~\Searth)$ & $T_{\rm HZ} (0.9~\Searth)$ & $T_{\rm HZ} (1.5~\Searth)$ & $T_{\rm HZ} (0.9~\Searth)$ & $T_{\rm HZ} (1.5~\Searth)$ & $T_{\rm HZ} (0.9~\Searth)$ & $T_{\rm HZ} (1.5~\Searth)$ \\
\hline
T1-b 	& 0.01111	&\multicolumn{2}{|c|}{\textbf{12.76--13.18}}	& \multicolumn{2}{|c|}{\textbf{8.96--9.28}}	& \multicolumn{2}{|c|}{\textbf{3.61--3.73}} 	& \multicolumn{2}{|c|}{\textbf{418--422}}\\
T1-c 	& 0.01522	&\multicolumn{2}{|c|}{\textbf{9.19--9.53}}		& \multicolumn{2}{|c|}{\textbf{6.39--6.63}}	& \multicolumn{2}{|c|}{\textbf{2.60--2.69}} 	& \multicolumn{2}{|c|}{\textbf{345--348}}\\
T1-d 	& 0.022	&\textbf{6.56--6.78} 	& 3.70--3.93			& \textbf{4.85--5.01} 	& 2.69--2.85		& \textbf{2.67--2.73} 	& 1.34--1.40 		& \textbf{489--493} 	& 222--227\\
T1-d 	& 0.058	& 0.32--0.41		& 0.15--0.24			& 0.24--0.30		& 0.11--0.17		& 0.14--0.17		& 0.06--0.09		& 28--32			& 11--15\\
T1-d 	& 0.146	& $<0.01$			& $<0.002$			& $<0.01$			& $<0.001$		& $<0.0007$		& $<0.0007$		& $<1.4$			& $<0.14$\\
\hline
\end{tabular} 
\label{waterlost1} 
\end{table*}

We used the method described in Section \ref{enerlim}, using an efficiency $\epsilon$ based on the hydrodynamical simulations of Section~\ref{hydrocode}.
We assumed an Earth-like composition to compute the masses of the planets \citep{Fortney2007} and we calculated $r_\textrm{F}$ following the method given in Section \ref{result_stoichio} for the three different XUV luminosity assumptions and for the different planets of the system.
We assumed that the semi-major axes of the planets remain constant throughout the evolution.

Figure \ref{massloss_TRAPPIST_0089_bcddd_alum_tinit_3_EOH_newrf_lin} show the hydrogen loss for the planets of the system for the three different XUV-luminosity trends as a function of time.
Table \ref{waterlost1} summarizes the results.
For $L_{\rm X}/L_{{\rm bol}}=10^{-5}$, we considered that $L_{\rm XUV}  = 5~L_{\rm X}$, as in Section \ref{ener_lim_esc}.
However, for $L_{\rm X}/L_{{\rm bol}}=10^{-3.4}$ and $L_{\rm X}/L_{{\rm bol}}=10^{-3.7}$, we used the value of \citet{Wheatley2016}: $L_{\rm XUV}  = 1.78~L_{\rm X}$.
For $L_{\rm X}/L_{{\rm bol}}=10^{-5}$, we find that planet b lose less than 4~$EO_H$ and planet c lose less than 3~$EO_H$ at the age of the system.
However, considering a higher XUV flux, this limit goes up to 13.5~$EO_H$ for planet b and 9.5~$EO_H$ for planet c.
Unless those planets have a big water content, they are therefore likely to be desiccated.

For planet d, due to the high uncertainty on its orbit, we find that at worst it could lose almost 7~$EO_H$ for an orbital separation of 0.022~au (assuming $L_{\rm X}/L_{\rm bol} =10^{-3.4}$ and that the planet never reaches the HZ).
For $L_{\rm X}/L_{\rm bol} =10^{-5.0}$, a planet d at 0.022~au loses more hydrogen than planet c at late ages. 
This is due to a combination of the effect of gravity on the cross-over mass (planet d being bigger than planet c, its $r_\textrm{F}$ is smaller) and XUV-flux which is lower for planet d which means that the efficiency $\epsilon$ is bigger.

However, considering TRAPPIST-1d is on the more probable orbit, at 0.058~au, we find that it loses between 0.06 and 0.41 $EO_H$.
It loses much less than if it was at 0.022~au because it is much farther away and enters the HZ much earlier.
If TRAPPIST-1d is at 0.146, it loses less than 0.01~$EO_H$.

As Hydrogen escapes faster than Oxygen, mass loss results in an Oxygen build up in the atmosphere \citep[e.g.,][]{LugerBarnes2015}.
We estimate the O$_2$ pressure in the atmosphere of the different planets (see Table \ref{waterlost1}). 
The pressure of O$_2$ can be as high as 500~bar for a planet d at 0.022~au.
It is even higher than the O$_2$ pressure for the two inner planets, because as the XUV flux planet d receives is lower than for planets b and c, its $r_\textrm{F}$ is smaller and it therefore loses much less oxygen than the two inner planets.
For a planet d at 0.058~au up to 30 bar of O$_2$ can build up in the atmosphere by the time it reaches the HZ.

The orbit of TRAPPIST-1d is not well constrained, but there is a high probability that it is in the HZ. 
This calculation shows that there is a non-negligible probability that this planet was able to retain a high fraction of an eventual water reservoir of one Earth Ocean, which makes it a very interesting astrobiology target.

Additional measurements of \Tpp's X-ray luminosity are needed in order to establish whether the values of \citet{Wheatley2016} correspond to a flare or not (a discussion about the quantitative effect of flares can be found in Section \ref{effect_flares}).
Besides, what might give us a deeper insight of the escape mechanisms for \Tpp's planets will be the characterization of the atmospheres of the three planets. 
Indeed, \citet{Belu2013} showed that the atmosphere of the planets of TRAPPIST-1 could be characterizable with facilities like JWST. 
The observation of these planets could therefore bring us informations on water delivery during the formation processes and their capacity to retain water.


\section{Discussion}\label{discussion}
\subsection{Why this result likely overestimate the loss}\label{whynot}

In this section, we show that the thought process we performed in the previous section, both following the standard way of computing mass loss \citep[as in][]{BarnesHeller2013} and using simple radiation-hydrodynamics calculations \citep[as in][]{OwenAlvarez2016} may actually be overestimating the mass loss.

\begin{itemize}
\item The time of the disk dispersal we consider here might be too short for such low mass objects.
The evolution of disks around UCDs is not well constrained, however it is reasonable to assume they dissipate between 3~Myr and 10~Myr \citep[disks around low mass stars tend to have longer lifetimes, e.g.][]{Pascucci2009, Liu2015, Downes2015}. 
Disks around UCDs could very well dissipate at an age of 10~Myr. 
As young UCDs are brighter than old UCDs, a later dissipation of the disk would mean that a planet is exposed less time and to a weaker XUV radiation, meaning that the planet would lose less water than what was calculated in this work.

For example, a $1~\Mearth$ planet orbiting a $0.04~\Msun$ BD at $0.013$~au would only lose $ 0.41~EO_H$ by the time it reaches the HZ if the disk dissipates at 10~Myr (instead of $0.48~EO_H$ if the disk dissipates at 3~Myr, see Figure \ref{massloss_Mp_alum5cst_Mbd_legend}, for $L_{\rm X}/L_{{\rm bol}} < 10^{-5}$).
And a planet orbiting a $0.06~\Msun$ BD at $0.013$~au would lose only $1.06~EO_H$ if the disk dissipates at 10~Myr (instead of $1.15~EO_H$ if the disk dissipates at 3~Myr).
Under this assumption of a longer lived protoplanetary disk, planets can therefore keep slightly more water.

	\begin{figure}
	\begin{center}
	\includegraphics[width=\linewidth]{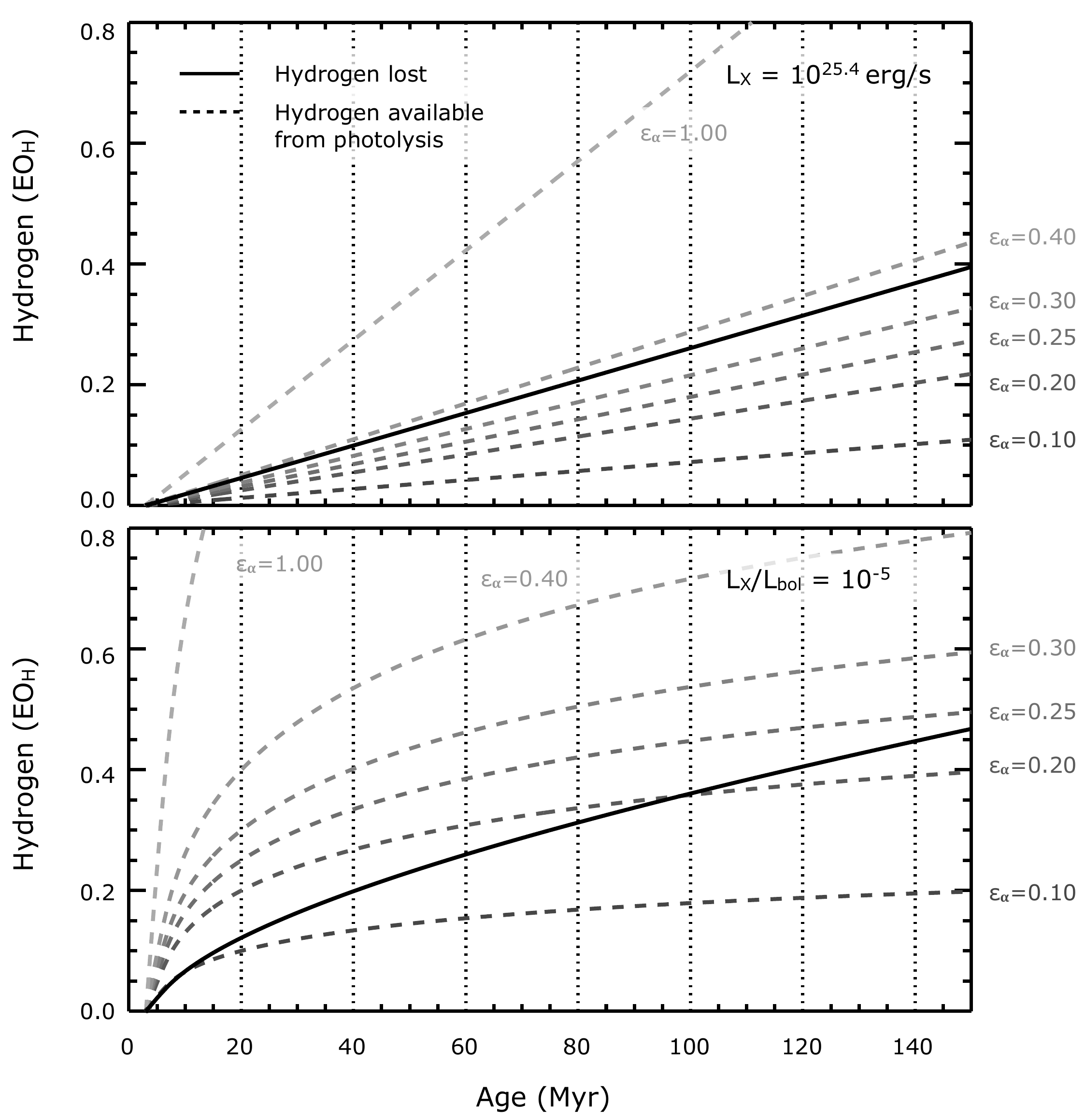}
	\caption{Hydrogen loss (as calculated in Section \ref{ener_lim_esc}) and hydrogen created by photolysis (dashed lines) as a function of time for an Earth-mass planet orbiting a $0.04~\Msun$ BD at 0.01~au. The top panel corresponds to $L_{\rm X} = 10^{25.4}$~erg/s and the bottom panel to $L_{\rm X}/L_{{\rm bol}}=10^{-5}$. The hydrogen quantity created by photolysis was calculated assuming a different efficiency of the photolysis process $\epsilon_\alpha$.}
	\label{Hloss_Hphoto_sma001_Mp_alum5_254ergs_Mbd_eps01_X5_tinit_1d6_EOH_rf_027}
	\end{center}
	\end{figure}

\item Water vapor photolysis is required to feed the loss in hydrogen atoms and is produced by FUV radiation (100--200 nm). 
UCD are too cool to produce a significant photospheric FUV flux and H$_2$O-photolysing radiation is likely to be restricted to the Lyman-alpha emission. 
Recent observations of 11 M-dwarfs by \citet{France2016} showed that the estimated energy flux in the Lyman-alpha band is equal to the flux in the XUV range.
Using this constraint, we can estimate the quantity of hydrogen produced by photodissociation.
Figure \ref{Hloss_Hphoto_sma001_Mp_alum5_254ergs_Mbd_eps01_X5_tinit_1d6_EOH_rf_027} shows the quantity of hydrogen lost according to our calculations of Section \ref{ener_lim_esc} and the quantity of hydrogen available.

If all the incoming FUV photons do photolyse H$_2$O molecules with a 100\% efficiency ($\epsilon_\alpha = 1$) and all the resulting hydrogen atoms remain available for the escape process then photolysis does not appear to be limiting the loss process. 
The production rate of hydrogen atoms exceeds the computed thermal loss assuming an hydrogen and oxygen mixture. 
In reality, however, only a fraction of the incoming FUV actually results in the loss of an hydrogen atom. 
Part of the incoming photons are absorbed by other compounds (in particular hydrogen in the Lyman alpha line) or backscattered to space. 
Then, products of H$_2$O photolysis (mainly OH and H) recombine through various chemical pathways. 
If the efficiency $\epsilon_\alpha$ is less than about 23\% then the loss rate becomes photolysis-limited. 
Although efficiency calculations would require detailed FUV radiative transfer and photochemical schemes, we can safely argue that efficiencies much lower than 23\% can be expected.

It is important to stress that the loss is likely to be photolysis-limited, which would allow us to calculate upper limits of the loss without the need of complex thermal and non-thermal escape models. 
At this point the FUV flux, which is the key input for H$_2$O photolysis, is only estimated based on the XUV/FUV ratio measured on earlier-type stars. 
Measuring the FUV of UCDs could allow us to put strong constraints on the water erosion on their planets.

\item The XUV flux considered here might be much higher than what is really emitted by UCDs.
Indeed, all Chandra observations of X-ray emissions of low mass objects \citep[e.g.][]{Berger2010, Williams2014} are actually non-detection for the UCD range.
New estimations from \citet{Osten2015} show that found upper limits for the X-ray luminosity for the object Luhman~16AB (WISE J104915.57--531906.1, L7.5 and T0.5 spectral types) are lower than what we used in this study: $L_{\rm X} < 10^{23}$~erg/s or $L_{\rm X}/L_{{\rm bol}} < 10^{-5.7}$.
Besides, \citet{Mohanty2002} show that in BDs' cool atmospheres the degree of ionization is very low so it is very possible that the mechanisms needed to emit X-rays are not efficient enough to produce the fluxes considered in this work. 
In which case, the computed mass loss would be lower than what we calculated here.
%

\item We find that the planets might lose several $EO_H$. Therefore, depending on the initial water content, some of them are in danger to be desiccated.
The close-in planets we consider may not have formed in situ but migrated from the outer regions of the disk and could have accumulated a large amount of water.
Both on the observational and theoretical side, it has been shown that planets could even have a large portion of their bulk made out of water \citep[Ocean planets, hypothesized by][]{Kuchner2003, Leger2004, Ogihara2009}. 
For example, \citet{Kaltenegger2013} have identified Kepler-62e and 62f to be possible ocean planets.
Furthermore, as discussed in \citet{Marty2012}, a large part of water can be trapped in the mantle to be released by geological events during the evolution of the planet, allowing a replenishment of the surface water content.


\item \citet{Johnson2013} showed using molecular-kinetic simulations that the mass loss saturates for high incoming energy, which could mean that in our case the mass loss would be smaller than what we calculated.
However, their results should be applied to our specific problem in order to be sure.
\end{itemize}

\subsection{Why this result is different from previous ones}\label{diff_res}

Unlike \citet{BarnesHeller2013}, we find that planets in the HZs of UCDs should in most cases be able to retain a non negligible portion of their initial reservoir of water. 
The main differences between \citet{BarnesHeller2013} and our work are the following. 
First, they used for the XUV radiation an observed upper limit for early-type M-dwarfs \citep[][which was the only available study at the time]{Pizzolato2003} for which the XUV luminosity scales as $10^{-3}$ times the bolometric luminosity.
We used here more recent estimates of X-ray emission of later-type dwarfs, which show that UCDs emit much less X-rays than earlier-type M-dwarfs \citep{Berger2010, Williams2014, Osten2015}. 
Second, in addition to the standard method used in \citet{BarnesHeller2013} and \citet{LugerBarnes2015}, we also improved the robustness of our results obtained with an improved energy-limited escape formalism using a better estimate of the fraction of the incoming energy that is transferred into gravitational energy through the mass loss ($\epsilon$) obtained with 1D radiation-hydrodynamic mass-loss simulations \citep{OwenAlvarez2016}.
Third, \citet{BarnesHeller2013} used a larger XUV cross-section for the planets.
However, using our hydrodynamical model, we found that for such small planets the XUV cross-section is very similar to the radius of the planet and only causes a difference of a few percents in the quantity of water lost.
Fourth, they considered a loss of only hydrogen atoms, while in this work we estimated the ratio of the escape fluxes of both hydrogen and oxygen atoms ($r_\textrm{F}$).
We found that in most of the configurations considered here, oxygen atoms are dragged away by the escaping hydrogen atoms, which is more favorable for water retention.

\subsection{The effect of flares}\label{effect_flares}

We only consider here quiescent energetic emissions. 
However BDs could emit energetic flares for a significant fraction of their lifetime in the H$\alpha$ emission line and in the U-band \citep{Schmidt2014a, Schmidt2014b}. 
This would also endanger the survival of a water reservoir. 
\citet{Gizis2013} showed that these flares can be as frequent as 1-2 times a month (e.g., the L1 dwarf W1906+40). 
W1906+40 experienced a white flare during $\sim2$ hours, which released an energy of $10^{31}$~erg (in a band $400$--$900$~nm). 
Let us consider here that the flare released in the XUV range 20\% of the energy it released in the $400$--$900$~nm band \citep[this proportion has been measured for Sun flares in][]{Kretzschmar2011}. 
This flare would then correspond to an energy $\sim 11$ times what we considered for the quiescent emission in the case of the constant XUV emission ($L_{\rm X} = 10^{25.4}$~erg/s). 
Using the Equation \ref{eqmassloss2}, we find that if such a flare happened during 2 hours every months (as could be the case for W1906+40), would reduce slightly the lifetime of the water reservoir. 
For example, a $1~\Mearth$ planet at $0.01$~au orbiting a UCD of $\Mbd = 0.04~\Msun$ would lose $0.191~EO_H$ instead of $0.189~EO_H$ before reaching the HZ at $\sim 60~$Myr (assuming $L_{\rm X} = 10^{25.4}$~erg/s, blue squares on Figure \ref{massloss_Mp_alum5cst_Mbd_legend}a).
A $1~\Mearth$ planet at $0.01$~au orbiting a UCD of $\Mbd = 0.08~\Msun$ would lose $0.899~EO_H$ instead of $0.890~EO_H$ before reaching the HZ at $\sim 300~$Myr 
Taking into account the flares therefore does not significantly change the results.

\subsection{Water retention does not equal habitability}

Water retention is not synonymous with habitability. 
Given that BD's HZs are very close-in, HZ planets feel strong tidal forces. 
This may affect their ability to host surface liquid water. 
For example, a lone planet would likely be in synchronous rotation. 
One can imagine that all liquid water might condense onto the night side (cold trap). 
However, this can be avoided if the atmosphere is dense enough to efficient redistribute heat \citep[e.g.,][]{Leconte2013}. 
In multiple-planet systems, a HZ planet's eccentricity can be excited and lead to significant tidal heating \citep{Barnes2009, Barnes2010, Bolmont2013, Bolmont2014sf2a}. 
In some cases tidal heating could trigger a runaway greenhouse state \citep{BarnesHeller2013}. 
In other situations, such as in the outer parts of the HZ or even exterior to the HZ, tidal heating may be beneficial by providing an additional source of heating and perhaps even by helping to drive plate tectonics \citep{Barnes2009}.    

\section{Conclusions}
\label{Conclusions}


Considering a very unfavorable scenario for water retention -- complete dissociation of water molecules -- and assuming different values for the X-ray luminosity of the UCDs, we find regions of parameter space (mass of UCD vs orbital distance of the planets) for which planets lose less than a few $EO_H$ (here equal to the hydrogen reservoir in one Earth Ocean) before reaching the HZ and can also spend a long time in the HZ.
When reaching the HZ, the remaining hydrogen can recombine with the remaining oxygen to form water molecules that can then condense.
The longer the planet spends in the HZ, the more time life has to eventually appear, evolve and be observable.
\citet{Bolmont2011} showed that the more massive the BD, the longer a close-in planet spends in the HZ. 
The low-mass BDs will always suffer from the fact that they cool down very fast and that at best, planets spend a few $100$~Myr in the HZ. 
Even though this could be enough time for life to appear, its potential detectability would be a rare event. 

This work therefore shows that there is a potential sweet spot for life around UCDs: planets between $0.01$~au and $0.04$~au orbiting BDs of masses between $\sim0.04~\Msun$ and $0.08~\Msun$ (assuming $L_{\rm X}/L_{{\rm bol}} =10^{-5}$ or $L_{\rm X} = 10^{25.4}$~erg/s) lose less than $1~EO_H$ while in runaway AND then spend a long time in the HZ ($\geq 1~$Gyr, according to \citealt{Bolmont2011}). 
Considering a higher X-ray luminosity ($L_{\rm X}/L_{{\rm bol}} =10^{-4.5}$ or $L_{\rm X} = 10^{26}$~erg/s), this sweet spot shifts towards higher orbital distances and higher UCD masses: planets between $\sim0.02$~au and $0.04$~au orbiting UCDs of mass between $\sim 0.06~\Msun$ and $0.08~\Msun$ lose less than $1~EO_H$ while in runaway AND then spend a long time in the HZ.
Of course, if one of the mechanisms considered here does not take place, or if the real XUV flux of BDs is lower than the upper value we considered, as we discussed in the previous Section \ref{whynot}, the sweet spot for life could widen towards the smaller orbital distances. 



We also investigated hydrogen losses in the Trappist-1 system \citep{Gillon2016}.
Assuming a X-ray quiescent emission comparable to a similar star as TRAPPIST-1 (VB~10), we find that the two inner planets of the system TRAPPIST-1 \citep{Gillon2016} might have lost up to $4~EO_H$ but that the third planet have lost less than $3~EO_H$.
Assuming a X-ray quiescent emission as high as \citet{Wheatley2016}, we find that if planet d has an orbital distance of 0.058~au (the most probably one from \citealt{Gillon2016}) it would have only lost at worst $\sim0.40~EO_H$.
If the X-ray luminosity of TRAPPIST-1 measured by \citet{Wheatley2016} is confirmed, the observation of the presence of water on the two inner planets would actually bring us informations about the planets initial water content.

Despite the lack of knowledge about escape mechanisms, in particular about the way hydrogen and oxygen jointly escape (or not), we find that there are possibilities that planets around UCDs might arrive in the HZ with an important water reservoir even without invoking an initial water reservoir larger than Earth's one.
As shown in Section \ref{whynot}, there might be even more possibilities if the loss of hydrogen is photolysis-limited, which would happen if the efficiency of this process is below 20\%.
Furthermore, planets in the HZs of BDs may be easy to detect in transit due to their large transit depths and short orbital periods \citep[at least for sufficiently bright sources;][]{Belu2013, Triaud2013}. 
Given their large abundance in the Solar neighborhood ($\sim$1300 have been detected to date; see \url{http://DwarfArchives.org}), such planets may be among the best nearby targets for atmospheric characterization with the JWST. 
In particular, the planets of TRAPPIST are an ideal laboratory to test the mechanisms of mass loss.


\section*{Acknowledgments}
The authors would like to thank Rory Barnes and Ren\'e Heller for bringing this subject to their attention.
The authors would also like to thank Rodrigo Luger for useful comments and for helping improve our manuscript.
The authors also thank the referee for the useful comments on the manuscript.

E. B. acknowledges that this work is part of the F.R.S.-FNRS ÓExtraOrDynHaÓ research project.
The work of E. B. was supported by the Hubert Curien Tournesol Program

I. R. acknowledges support from the Spanish Ministry of Economy and Competitiveness (MINECO) through grant ESP2014-57495-C2-2-R.

F. S. acknowledges support from the Programme National de Plan\'etologie (PNP). 
S. N. R. thanks the Agence Nationale pour la Recherche for support via grant ANR-13-BS05-0003-002 (project MOJO).

J. E. O. acknowledges support by NASA through Hubble Fellowship grant HST-HF2-51346.001-A awarded by the Space Telescope Science Institute, which is operated by the Association of Universities for Research in Astronomy, Inc., for NASA, under contract NAS 5-26555.

M. G. is Research Associate at the F.R.S.-FNRS.

\bsp

\label{lastpage}

\end{document}